\pdfoutput=1
\documentclass[12pt]{article}

\usepackage[utf8]{inputenc}
\usepackage{graphicx,psfrag,epsf,color}
\usepackage{amsmath,amssymb,amsfonts}
\usepackage{hyperref}
\usepackage{array}
\usepackage{bm} 
\usepackage{cite}
\usepackage{hyperref}
\usepackage{multirow}
\usepackage{rotating}
\usepackage{slashed}
\usepackage{textcomp}
\usepackage{float}
\newcommand{\lessim}{\mbox{\raisebox{-3pt}{$\stackrel{<}{\sim}$}}}

\newcommand{\be}{\begin{equation}}
\newcommand{\ee}{\end{equation}}
\newcommand{\bea}{\begin{eqnarray}}
\newcommand{\eea}{\end{eqnarray}}
\newcommand{\bi}{\begin{itemize}}
\newcommand{\ei}{\end{itemize}}
\newcommand{\ben}{\begin{enumerate}}
\newcommand{\een}{\end{enumerate}}
\newcommand{\bt}{\begin{tabular}}
\newcommand{\et}{\end{tabular}}

\newcommand{\Eres}{E^\gamma_{\rm res}}

\newcommand{\nn}{\nonumber}
\newcommand{\mDM}{m_\chi}
\newcommand{\muh}{\mu_h}
\newcommand{\mus}{\mu_s}
\newcommand{\muj}{\mu_j}
\newcommand{\nuh}{\nu_h}
\newcommand{\nus}{\nu_s}
\newcommand{\NLL}{\text{NLL}}

\numberwithin{equation}{section}

\newcounter{SLQ}

\newcounter{MBQ}

\begin{document}
\allowdisplaybreaks

\begin{titlepage}

\begin{flushright}
{\small
TUM-HEP-1426/22\\
November 24, 2022
}
\end{flushright}

\vskip1cm
\begin{center}
{\Large \bf 
Electroweak resummation of neutralino 
dark-matter\\[0.0cm]  annihilation
into high-energy photons \\[0.2cm]}
\end{center}

\vspace{0.45cm}
\begin{center}
{\sc M.~Beneke$^{a}$, S.~Lederer$^{a}$,} and {\sc C. Peset$^{b}$}\\[6mm]
{\it ${}^a$Physik Department T31,\\
James-Franck-Stra\ss e~1, 
Technische Universit\"at M\"unchen,\\
D--85748 Garching, Germany}\\[0.3cm]
{\it ${}^b$Dpto. de F{\'i}sica Te{\'o}rica \& 
IPARCOS, \\
Universidad Complutense de Madrid, \\
E--28040 Madrid, Spain}
\end{center}

\vspace{0.55cm}
\begin{abstract}
\vskip0.2cm\noindent
We consider the resummation of large electroweak Sudakov logarithms 
for the annihilation of neutralino DM with $\mathcal{O}$(TeV) mass 
to high-energy photons in the minimal 
supersymmetric standard model, extending previous work on 
the minimal wino and Higgsino models. We find that NLL 
resummation reduces the yield of photons by about 
$20\%$ for Higgsino-dominated DM at masses around 1~TeV,  
and up to  $45\%$ for neutralinos with larger wino 
admixture at heavier masses near 3~TeV. This sizable effect is relevant when 
observations or exclusion limits are translated into MSSM 
parameter-space constraints.  
\end{abstract}
\end{titlepage}


\section{Introduction}
\label{sec:intro}

Dark matter (DM), for which solid cosmological \cite{Ade:2015xua} and astrophysical \cite{Rubin:1980zd} observations exist, is the most direct indication of physics beyond the Standard Model, yet its fundamental nature is unknown.
Among the many DM candidates, weakly interacting massive particles (WIMPs) are well motivated as they provide the observed DM relic abundance naturally \cite{Steigman:2012nb,Arcadi:2017kky}, are thermalized in the early universe, and may be 
related to the electroweak scale.

A well-known WIMP is the neutralino, the lightest supersymmetric particle (LSP) of the (R-parity preserving) Minimal Supersymmetric Standard Model (MSSM) or its extensions. It can be a generic admixture of the bino, wino and Higgsino gauge multiplets. The relic abundance of the neutralino and its other signatures \cite{Jungman:1995df,ArkaniHamed:2006mb} depend on the exact composition. 
Neutralinos may undergo co-annihilation with other nearly degenerate states from gauge multiplets and, if their mass $m_\chi$ is larger than approximately 1~TeV, have Sommerfeld-enhanced annihilation cross sections \cite{Hisano:2004ds,Hryczuk:2010zi,Beneke:2014hja,Beneke:2016ync}. Both effects are crucial for precisely determining the DM relic
abundance as well as gamma-ray signals from neutralino annihilation in the present universe.

Null results from direct and indirect searches \cite{Arcadi:2017kky,Roszkowski:2017nbc} of WIMPs at the electroweak mass-scale have pushed the interest towards DM candidates with 
$\mathcal{O}(\mbox{TeV})$ masses. 
In the mass range of $\mDM \sim 1-3$~TeV, there are still several regions of the MSSM parameter space that have not been ruled out by experimental searches \cite{Hryczuk:2019nql,Athron:2018vxy,Kvellestad:2019vxm,Beneke:2016jpw}. Such large masses cannot be excluded by ongoing direct detection or collider searches and the best way to probe them is through the products of their annihilation. WIMPs move at non-relativistic velocities $v\sim 10^{-3}$ in the Milky Way halo. As a consequence, the high-energy ($E_\gamma\sim m_\chi$) photon yield of their annihilation will be in the form of almost monochromatic gamma-rays, easily distinguishable from other astrophysical backgrounds. This distinctive signal has already been searched for in the Galactic Center by experiments with  large  effective  area  at TeV energies, such as Fermi-LAT \cite{Ackermann:2015lka}, H.E.S.S. \cite{HESS:2018cbt}, VERITAS \cite{Archambault:2017wyh} or MAGIC \cite{Aleksic:2013xea}. Projections for the next-generation Cherenkov telescope array (CTA) \cite{CTAConsortium:2018tzg}, which is expected to start taking data within the next decade, predict improvements on the existing limits on spectral lines by one order of magnitude for TeV-scale gamma-ray energies. For WIMP DM in this mass range, given the apparatus' energy resolution of $\mathcal{O}(10\%)$ \cite{CTAO_2021_5499840} 
and given that only one photon per annihilation can be observed on earth, the proper observable is the semi-inclusive annihilation process $\chi^0\chi^0\to \gamma + X$ \cite{Baumgart:2017nsr,Beneke:2018ssm}, where $X$ is a collection of unobserved annihilation products, rather than the annihilation into two photons and $\gamma Z$ as is usually assumed in these searches. 

For neutralino DM, the theoretical prediction of spectral-line signals and the semi-inclusive photon spectrum near maximal photon energy has significant non-perturbative effects. The expansion in the electroweak gauge couplings breaks down for two different reasons: first, the almost back-to-back annihilation of neutralinos into a gamma-ray with energy $E_\gamma\sim \mDM$ and an electroweak ``jet'' $X$ produces large Sudakov logarithms of the ratio of the electroweak and DM  scales that must be resummed \cite{Baumgart:2014vma,Ovanesyan:2014fwa,Bauer:2014ula}.  Additionally, the electroweak Yukawa force between the neutralinos and charginos, the charged mass eigenstates, implies that an infinite number of ladder diagrams contribute at the same order, and must also be summed. This phenomenon is known as Sommerfeld enhancement and has been thoroughly studied for WIMPs \cite{Hisano:2003ec,Hisano:2004ds,ArkaniHamed:2008qn} and for the MSSM \cite{Hryczuk:2010zi,Beneke:2014hja,Beneke:2016ync,Beneke:2016jpw}. 
Both of these effects are efficiently handled by the use of effective field theories (EFTs), in particular the combined use of non-relativistic (NR) EFT \cite{Beneke:2012tg,Beneke:2014gja,Hellmann:2013jxa} to describe the heavy particles prior to annihilation and soft-collinear effective theory (SCET) \cite{Bauer:2000yr,Bauer:2001yt,Beneke:2002ph,Beneke:2002ni} to describe the energetic annihilation products. This approach has been used in recent works to describe the cases of pure wino \cite{Baumgart:2014vma,Ovanesyan:2014fwa,Bauer:2014ula,Ovanesyan:2016vkk,Baumgart:2017nsr,Beneke:2018ssm,Beneke:2019vhz,Beneke:2019qaa} and pure Higgsino \cite{Beneke:2019gtg}.
Of special relevance to us is \cite{Beneke:2019vhz} where all the functions for the evaluation of the factorization formulas at next-to-leading-logarithmic (NLL) order are provided for the experimental photon-energy resolution in the narrow $\Eres\sim m_W^2/\mDM$ or intermediate $\Eres\sim m_W$ regime, where $m_W$ denotes the mass of the W-boson. For TeV-energy photons, the intermediate-energy regime is the most appropriate one for current instruments.

While the Sommerfeld effect can result in orders-of-magnitude changes of the neutralino annihilation cross section near specific, resonant mass values, the above-mentioned work for pure wino and Higgsino showed that the summation of electroweak Sudakov logarithms typically leads to a reduction of the photon yield by up to a factor of two for TeV mass neutralinos. Such sizable effects are important when observations or exclusion limits are translated into model parameter constraints or compared with constraints from other observations. In this paper we compute for the first time the semi-inclusive neutralino annihilation into photons for general neutralinos, as they appear in the MSSM, which are mixed states of the wino, Higgsino and bino, with NLL accuracy for intermediate experimental resolution. The necessary EFT framework is introduced in section \ref{sec:EFT} and the anomalous dimensions of the factorized photon spectrum are given in section \ref{sec:Ann}. We apply this calculation to a series of numerical benchmark points for TeV-scale DM and present the results in section \ref{sec:results}. The main text is supplemented by technical appendices, 
covering electroweak symmetry breaking (EWSB) in the non-relativistic 
MSSM, details on the computation of the hard anomalous dimension, 
the resummed functions in the factorization formula, and the treatment of the partially rather than fully degenerate 
neutralino sector.


\section{EFT framework for neutralino dark-matter annihilation}
\label{sec:EFT}

Throughout this section we closely follow the framework of \cite{Beneke:2019vhz,Beneke:2019gtg}, to which we refer for further details on notation and definitions in the non-relativistic and soft-collinear effective theory. Our MSSM notation and conventions coincide with \cite{Rosiek:1995kg} except for the gaugino fields which we rotate by a factor $i$ with respect to that reference. 

Given the current experimental bounds on supersymmetric particle masses \cite{CMS:2021edw, ATLAS:2022rcw, Workman:2022ynf} and CP violation, it is reasonable to consider the CP-symmetry preserving MSSM with all supersymmetric particle masses above 1~TeV, in which case electroweak and non-relativistic resummation is necessary since $m_\chi\gg m_W$. Our framework is especially relevant in the case of a heavily mixed neutralino state. In particular, we consider models where the SU(2)$\times$U(1)$_Y$ gauge boson and Higgs superpartners are the lightest while all others are heavier and are not relevant to co-annihilation processes. We further consider the two possibilities that the additional Higgs doublet in the MSSM is either  heavy (mass of order of the neutralino or 
heavier), or light (mass of order of the Standard Model Higgs boson), and distinguish them by the parameter $n_H=1,2$, respectively.  The Lagrangian of this model is given by
\begin{align}
\mathcal{L}=\mathcal{L}_\text{SM}+(n_H-1)\mathcal{L}_H+\mathcal{L}_\chi,
\end{align}
where $\mathcal{L}_\text{SM}$ contains all the interactions of Standard Model (SM) fields, $\mathcal{L}_H$ their interactions with the extra Higgs doublet and $\mathcal{L}_\chi$ is the NR Lagrangian encoding the kinetic terms and interactions of the gaugino and Higgsino fields. At leading order in the NR expansion, $\mathcal{L}_\chi$ is given by 
\begin{align}
\label{LagNRg}
\mathcal{L}_\chi&=\xi_A^\dagger\left(i D^0+ \frac{{\bf D}^2}{2\mDM}-\delta m_2\right)\xi_A+\xi_B^\dagger\left(i \partial^0+ \frac{{\boldsymbol{\partial}}^2}{2\mDM}-\delta m_1\right)\xi_B\nn\\
&+\eta_{H}^{\dagger}\left(iD^0+ \frac{{\bf D}^2}{2\mDM}-\delta m_H\right)\eta_H+\zeta_H^{c\dagger}\left(i D^0-\frac{{\bf D}^2}{2\mDM}+\delta m_H\right)\zeta_H^c\nn\\
&+ (H_+^T \epsilon) \left[g_2T^a\left(\xi_A^{a\,\dagger}\eta_H-\xi_A^{a\,c\dagger}\zeta^c_H\right)-\frac{g_1}{2}\left(\xi_B^\dagger\eta_H-\xi_B^{c\dagger}\zeta^c_H\right)\right]\nn\\
&- \left[g_2\left(\eta_H^\dagger\xi_A^a-\zeta_H^{c\dagger}\xi_A^{ac}\right)T^a-\frac{g_1}{2}\left(\eta_H^\dagger\xi_B-\zeta_H^{c\dagger}\xi_B^{c}\right)\right] (\epsilon H_+^*)\,.
\end{align}
with $\delta m_{1,2}=M_{1,2}- m_\chi$ and $\delta m_H = |\mu|- m_\chi$, and $ m_\chi=\min(M_1,M_2,|\mu|)$ in this context. The NR fermionic bino, wino and (anti-) Higgsino fields $\xi_B, \,\xi_A$ and $\eta_H,\,\zeta_H$ are two-component spinors with SU(2)$\times$U(1)$_Y$ charges $\bf 1_0$, $\bf 3_0$ , $\bf 2_{-1/2}$ and, $\bf \bar{2}_{1/2}$, respectively. Charge-conjugated fields obtain the superscript 
``c'' and $\epsilon=i \sigma^2$ operates on the SU(2) index of the 
doublet $H_+$. The convention for the SU(2)$\times$U(1)$_Y$ covariant derivative is ${iD_\mu \equiv i\partial_\mu - g_1YB_\mu - g_2T^aA^a_\mu}$.
$H_+$ denotes a linear combination of the neutral scalar Higgs boson mass eigenstates $h$ and~$H$,
\begin{align}
H_+ &=
\label{def:Hpm2text}
\frac{s_{\alpha_H}+\mathrm{sign}(\mu) c_{\alpha_H}}{\sqrt{2}} \, h +
(n_H-1)\frac{c_{\alpha_H}-\mathrm{sign}(\mu) s_{\alpha_H}}{\sqrt{2}} \, H 
\,,
\end{align}
which depends on the Higgs mixing angle $\alpha_H$ before EWSB. ($ s_{\alpha_H}$ and $c_{\alpha_H}$ denote the sin and cos of 
the angle.)
Details on the Higgs interactions and EWSB are given in 
appendix~\ref{app:EWSB}. 
In \eqref{LagNRg} we are implicitly assuming that 
$\delta m_X$ is at most of order of the electroweak 
scale $m_W$, which represents the infrared scale for 
the Sudakov logarithms. That is, all neutralinos and gauginos are degenerate for the purpose of Sudakov resummation, and therefore present in the NR effective theory. We discuss the extension to non-degenerate models in section \ref{sec:nondegenerate}. 

Being defined at the NREFT matching scale $\mathcal{O}(2\mDM)$, the fermions in \eqref{LagNRg} represent gauge multiplets before EWSB.
After EWSB, the NR gauginos and Higgsinos are combined into new mass eigenstates, which produce four neutralinos and two charginos. Different from the NRMSSM construction in \cite{Beneke:2012tg}, in \eqref{LagNRg} we take the NR limit before EWSB because we will sum electroweak logarithms by renormalization-group evolution from the hard scale $2\mDM\gg m_W$, which does not know about EWSB, to the EWSB scale $m_W$. 
By explicit computation of the chargino and neutralino masses and mixing matrices, we checked that these two limits are interchangeable up to small relative power corrections of order $m_W/m_\chi$, 
which are beyond the accuracy of the effective Lagrangians. After rotation of the mass matrix in the NR unbroken gauge basis to the mass eigenstate basis after EWSB, 
one linear combination of the neutral Higgsino fields decouples from the gauginos in the mass matrix, see (\ref{MNNR}). Additionally, we find that in the limit $|M_1-M_2| \rightarrow 0$ one linear gaugino combination decouples from the Higgsino fields in the mass matrix. Further details on the computation of the mass matrices in the NRMSSM can be found in appendix~\ref{app:EWSB}. 

The hard neutralino pair-annihilation process in the NREFT-SCET framework is described by the effective Lagrangian \cite{Beneke:2019vhz}
\begin{equation}
\label{def:Lann}
\mathcal{L}_\text{ann}=\frac{1}{2m_\chi}\sum_{i=1}^9\int ds dt \,\hat C_i(s,t,\mu)\mathcal{O}_i(s,t),
\end{equation}  
where $\hat C_i$ are the Wilson coefficients evolved to the resummation scale $\mu$ (not to be confused with the Higgsino mass parameter) and the operators take the form 
\begin{align}
\mathcal{O}_i=\chi^{c\dagger}\Gamma^{\mu\nu}T^{AB}_i\chi \,
\mathcal{A}^A_{\perp c,\mu}(sn_+)\mathcal{A}^B_{\perp\bar{c},\nu}(tn_-).
\label{def:genOp}
\end{align}
Here 
\begin{align}
\chi=\begin{pmatrix}
\xi_B&\xi_A^1&\xi_A^2&\xi_A^3&\zeta_H^1&\zeta_H^2&\eta_H^1&\eta_H^2
\end{pmatrix}^T
\label{def:chi}
\end{align} 
denotes the vector of SU(2)$\times$U(1)$_Y$ eigenstate two-component non-relativistic spinors, and $\mathcal{A}^A_{\perp,\mu}$ is the (anti-)collinear gauge boson field including light-like Wilson lines, with index value $A=1,2,3$ corresponding to the SU(2), and $A=4$ to the U(1)$_Y$ group \cite{Beneke:2019gtg}. For clarity we use $\mathcal{A}^4=\mathcal{B}$ below so $T^{C}$ are the usual SU(2) generators in the respective adjoint or fundamental representation of the wino or Higgsino.\footnote{
We usually leave the gauge indices on the fermion fields implicit, except for operator $\mathcal{O}_6$, where the wino adjoint index $C$ must be contracted explicitly with the gauge boson index.
} 
We find that in the MSSM the following nine operators 
contribute to neutralino semi-inclusive annihilation into photons:
\begin{align}\label{def:ops}
\mathcal{O}_1&=\xi_A^{c\dagger}\Gamma^{\mu\nu}\xi_A
\mathcal{A}^C_{\perp c,\mu}(sn_+) \mathcal{A}^C_{\perp \bar{c},\nu}(tn_-)\,,\nn\\
\mathcal{O}_2&=\frac12 \,\xi_A^{c\dagger}\Gamma^{\mu\nu}
\{T^C,T^D\}\xi_A\,
\mathcal{A}^C_{\perp c,\mu}(sn_+) \mathcal{A}^D_{\perp \bar{c},\nu}(tn_-)\,,\nn\\
\mathcal{O}_3&=\xi_B^{c\dagger}\Gamma^{\mu\nu}\xi_B\,
\mathcal{A}^C_{\perp c,\mu}(sn_+) \mathcal{A}^C_{\perp \bar{c},\nu}(tn_-)\,,\nn\\[0.15cm]
\mathcal{O}_4&=(\zeta_H^{c\dagger}\Gamma^{\mu\nu}\eta_H+\eta_H^{c\dagger}\Gamma^{\mu\nu}\zeta_H)\,
\mathcal{A}^C_{\perp c,\mu}(sn_+) \mathcal{A}^C_{\perp \bar{c},\nu}(tn_-)\,,\nn\\[0.15cm]
\mathcal{O}_5&=(\zeta_H^{c\dagger}\Gamma^{\mu\nu}T^C\eta_H + \eta_H^{c\dagger}\,\Gamma^{\mu\nu}(T^C)^T\zeta_H)\,
[\mathcal{A}^C_{\perp c,\mu}(sn_+) \mathcal{B}_{\perp \bar{c},\nu}(tn_-)+\mathcal{B}_{\perp c,\mu}(sn_+) \mathcal{A}^C_{\perp \bar{c},\nu}(tn_-)]\,,\nn\\
\mathcal{O}_6&=\frac{1}{2}(\xi_B^{c\dagger}\Gamma^{\mu\nu}\xi_A^C+\xi_A^{C\,c\dagger}\Gamma^{\mu\nu}\xi_B)[\mathcal{A}^C_{\perp c,\mu}(sn_+) \mathcal{B}_{\perp \bar{c},\nu}(tn_-)+\mathcal{B}_{\perp c,\mu}(sn_+) \mathcal{A}^C_{\perp \bar{c},\nu}(tn_-)]\,,\nn\\
\mathcal{O}_7&=\xi_A^{c\dagger}\Gamma^{\mu\nu}\xi_A\,
\mathcal{B}_{\perp c,\mu}(sn_+) \mathcal{B}_{\perp \bar{c},\nu}(tn_-)\,,\nn\\[0.15cm]
\mathcal{O}_{8}&=\xi_B^{c\dagger}\Gamma^{\mu\nu}\xi_B\,
\mathcal{B}_{\perp c,\mu}(sn_+) \mathcal{B}_{\perp \bar{c},\nu}(tn_-)\,,\nn\\[0.1cm]
\mathcal{O}_{9}&=(\zeta_H^{c\dagger}\Gamma^{\mu\nu}\eta_H+\eta_H^{c\dagger}\Gamma^{\mu\nu}\zeta_H)\,
\mathcal{B}_{\perp c,\mu}(sn_+) \mathcal{B}_{\perp \bar{c},\nu}(tn_-)\,.
\end{align}
We restrict ourselves to CP-conserving, spin-0 S-wave annihilation operators, hence the spin-matrix $\Gamma^{\mu\nu}$ in Weyl-spinor space is simply, ${\Gamma_{\mu\nu} = \mathbf{1}\cdot \frac{1}{2}\epsilon_{\mu\nu\rho\sigma}n_+^\rho n_-^\sigma}$ \cite{Beneke:2019vhz}. We only describe annihilation processes of heavy NR fields, hence there is no need for the hermitian conjugated operators, which would describe pair creation. We note that at leading power, the only relevant hard processes are annihilations into two gauge bosons \cite{Beneke:2019vhz}. A priori, the final state is allowed to contain Higgs bosons, however this is only possible via spin-1 operators, which cannot contribute to the annihilation process $\chi^0\chi^0 \to \gamma+X$ of an initial state of two identical fermions. Note that $\mathcal{O}_6$ contains wino and bino fields and is therefore not present in pure bino, wino or Higgsino models.

The NR effective Lagrangian \eqref{LagNRg} in the unbroken 
electroweak-symmetry phase together with 
the soft-collinear Lagrangian, which describes the dynamics 
of the (anti-)collinear gauge boson fields, is suitable 
for the calculation of the hard anomalous dimension of 
the above operators, which describes the evolution of 
the coefficient function from the high scale $m_\chi$ to 
the infrared scale $m_W$. The computation of the low-scale 
matrix elements, which includes the Sommerfeld effect 
on the annihilation cross section, however, uses the 
NRMSSM and SCET Lagrangian after EWSB. EWSB shifts the 
eigenvalues of the mass matrix by amounts up to 
$\mathcal{O}(m_W)$ and the physical masses after EWSB 
must be used in the Sommerfeld calculation. However, these 
mass shifts are irrelevant for the electroweak 
Sudakov resummation.


\section{$\chi^0\chi^0 \to \gamma+X$ at NLL}
\label{sec:Ann}

Reaching NLL precision on the resummation of large electroweak logarithms, requires 1) the initial conditions of the functions in the factorized annihilation rate at tree-level in the expansion in the U(1)$_Y$ and SU(2) couplings $\alpha_1,\alpha_2$ and at leading power in the mass ratio $m_W/\mDM$, and 2)  the anomalous dimensions in the renormalization-group equations (RGEs) at one-loop, and two-loop for the cusp pieces.
The contributions to the anomalous dimension coming from the exchange of gauge bosons are determined by the gauge representations of our fields, allowing us to extract the two-loop cusp pieces without explicit computation \cite{Beneke:2009rj}. However, the contributions from Higgs exchange between gauginos and Higgsinos must be 
computed separately and are a novel result of our work. Details on their computation are given in appendix \ref{app:ADM}. 
 
\subsection{Anomalous dimensions for the factorized annihilation rate}\label{sec:ADM}

The photon energy spectrum of DM annihilation $\chi^0\chi^0\to \gamma+X$ including the Sommerfeld effect and electroweak Sudakov resummation can be written in the factorized form
\begin{align}
\frac{d(\sigma v_\text{rel})_{\gamma X}}{d E_\gamma}=2\sum_{IJ}S_{IJ}\Gamma_{IJ}(E_\gamma)
\end{align}
where the states $I=\{i_1i_2\},J=\{j_1j_2\}$ span over a total of $14=10+4$ electrically neutral two-particle states, 10 for neutralinos and 4 for charginos.
The Sommerfeld factor $S_{IJ}$ is computed in the MSSM following \cite{Beneke:2014gja} and does not depend on the photon energy $E_\gamma$. The annihilation rate $\Gamma_{IJ}$ for intermediate resolution  $\Eres\sim m_W$ was obtained in \cite{Beneke:2019vhz,Beneke:2019gtg} and reads
\begin{eqnarray}
\Gamma_{IJ}(E_\gamma)&=&\frac{1}{(\sqrt{2})^{n_{id}}}\frac{1}{4}\frac{2}{\pi m_\chi}\sum_{i,j=1}^9H_{ij}(\muh,\mu)Z_\gamma^{YW}(\mus,\nuh,\mu,\nu)\nn\\
&&\times\int d\omega \left[J_{\text{int}}^{\text{SU(2)}}(p^2,\muj,\mu)W^{ij,\text{SU(2)}}_{IJ,WY}(\omega,\mus,\nus,\mu,\nu)\right.\nn\\
&&\left.\hspace*{1.5cm}+J_{\text{int}}^{\text{U(1)}}(p^2,\muj,\mu)W^{ij,\text{U(1)}}_{IJ,WY}(\omega,\mus,\nus,\mu,\nu)\right],
\label{def:GammaIJ}
\end{eqnarray}
where $p^2=4m_\chi(m_\chi-E_\gamma-\omega/2)$, $H_{ij}=C_iC_j^*$ is the hard function given by the Fourier-transformed (momentum-space) Wilson coefficients $C_i$, $Z_\gamma^{YW}$ is the photon (anti-collinear) jet function, $J_{\text{int}}^{U(1),\,SU(2)}$ are the hard-collinear functions already split into the separate gauge group contributions, and $W^{ij,U(1)}_{IJ,WY}$ and $W^{ij,SU(2)}_{IJ,WY}$ are the corresponding soft functions. 
In \eqref{def:GammaIJ} we write explicitly the dependence not only 
on the virtuality and rapidity factorization scales $\mu$ and $\nu$ 
but also on the natural matching scales for each function. We thus define the hard scales $\muh\sim\nuh\sim 2\mDM$, the jet scale $\muj\sim\sqrt{2\mDM m_W}$ and the soft scales $\mus\sim\nus\sim m_W$. We evolve all objects to a common scale $(\mu,\nu)$ using their respective RGEs. We explicitly checked that, for the two resummation schemes proposed in \cite{Beneke:2019vhz}, the residual dependence on the choice of $(\mu,\nu)$ in the NLL is completely negligible, and so for simplicity we choose from now on $(\mu,\nu)=(\mus,\nus)$. With this choice the soft function does not have to be evolved.

The semi-inclusive annihilation of neutralinos or charginos into $\gamma+X$ at NLL is computed for the primary photon energy spectrum, which is smeared with an instrument-specific resolution function of some width $\Eres$ in energy. To mimic this effect, we define
\begin{align}\label{NLLann}
\left\langle\sigma v_{rel} \right\rangle_{\gamma X}
\equiv 
\langle \sigma v_{rel}\rangle_{\gamma X} (\Eres)=\int_{m_\chi-\Eres}^{m_\chi}dE_\gamma
\frac{d(\sigma v_{rel})_{\gamma X}}{dE_\gamma}=2\sum_{IJ}S_{IJ}\Gamma^\NLL_{IJ}(\Eres),
\end{align}
where \eqref{NLLann} implicitly defines $\Gamma^\NLL_{IJ}(\Eres)$ as the integral of $\Gamma_{IJ}(E_\gamma)$. 

To NLL accuracy, we use that the tree-level soft function is 
of the form
\begin{align}
W^{ij}_{IJ,WY}(\omega,\mus,\nus,\mus,\nus)=w^{ij}_{IJ,WY}(\mu_s)\delta(\omega)
\end{align}
to simplify  \eqref{def:GammaIJ} to
\begin{align}\label{GammaSoft}
&\Gamma^\NLL_{IJ}(\mu_h,\mu_j,\mu_s,\nu_h,\nu_s,E^\gamma_\text{res})=\frac{1}{(\sqrt{2})^{n_{id}}}\frac{1}{2\pi m_\chi}\sum_{i,j=1}^9 H_{ij}(\mu_h,\mu_s)Z_\gamma^{YW}(\mu_s,\nu_h,\mu_s,\nu_s)\nn\\
&\times \frac{1}{4m_\chi}\left[U_J^\text{SU(2)}(\mu_j,\mu_s)\frac{e^{-\gamma_E\eta}}{\Gamma(1+\eta)}w^{ij,\text{SU(2)}}_{IJ,WY}(\mu_s)\left(\frac{4m_\chi E_\text{res}^\gamma}{\mu_j^2}\right)^\eta+U_J^\text{U(1)}(\mu_j,\mu_s)w^{ij,\text{U(1)}}_{IJ,WY}(\mu_s)\right],
\end{align}
where the definition of $\eta$ and the evolution factors $U_J$ can be found in \cite{Beneke:2019vhz,Beneke:2019gtg}.

The photon and unobserved jet functions depend only on the final annihilation products and we refer again to \cite{Beneke:2019vhz,Beneke:2019gtg} for their explicit form.\footnote{Due to a different Higgsino charge assignment there is a sign difference with respect to \cite{Beneke:2019gtg} in $Z_\gamma^{34}$ which is compensated by a sign difference in $C_5$ in \eqref{TLhard}.
} 
The soft function at tree level is fully determined by the mixing of gauginos and Higgsinos into neutralinos and charginos as detailed in appendix~\ref{app:EWSB} and \ref{app:FunctionsDef}. 

The resummed hard function is given by the solutions to the RGE
\begin{align}\label{eq:hardRGE}
\frac{d C_i}{d\ln\mu}=\gamma_{ij}^T C_j\,.
\end{align}
Tree-level matching of the Wilson coefficients at the hard scale results in 
\begin{align}\label{TLhard}
C_{1,3,6,7,8}=0 \quad \text{and}\quad C_2=g_2^2,\,\,C_4=\frac{g_2^2}{4},\,\,C_5 = -\frac{g_1g_2}{2},\,\,C_9=\frac{g_1^2}{4}\,.
\end{align}
NLL resummation further needs the one-loop anomalous dimension matrix $\gamma$. 
It can be split into three block-diagonal terms according to the SU(2) and U(1)$_{\rm Y}$ charges of the annihilation products such that
\begin{align}\label{def:Gamma}
\gamma=\begin{pmatrix}
\gamma_{AA}&&\\
&\gamma_{AB}&\\
&&\gamma_{BB}
\end{pmatrix}.
\end{align}
At the one-loop order we find 
\begin{align}
\gamma_{AA}=\begin{pmatrix}
\gamma_{AA}^{(1)} & 0 & 0 & \frac{\alpha_2}{4\pi}3a_+^2\\
\frac{\alpha_2}{4\pi}8(1-i\pi)&\gamma_{AA}^{(2)} & 0 & \frac{\alpha_2}{4\pi}2a_+^2\\
 0 & 0 & \gamma_{AA}^{(3)}&\frac{\alpha_1}{4\pi}a_+^2\\
\frac{\alpha_2}{4\pi}4a_+^2 & 0 & \frac{\alpha_1}{4\pi}4a_+^2&\gamma_{AA}^{(4)}
\end{pmatrix}\label{GammaAA},
\end{align}
\begin{align}
\gamma_{AB}&=\begin{pmatrix}
\gamma_{AB}^{(1)}&-\frac{\sqrt{\alpha_1\alpha_2}}{4\pi}4a_+^2\\[3pt]
-\frac{\sqrt{\alpha_1\alpha_2}}{4\pi}2a_+^2&\gamma_{AB}^{(2)}
\end{pmatrix},\label{GammaAB}
\end{align}
and
\begin{align}
\gamma_{BB}&=\begin{pmatrix}
\frac{\alpha_1}{4\pi}(-2\beta_{0,\text{U(1)}})+\frac{\alpha_2}{4\pi}4a_+^2 & 0 & \frac{\alpha_2}{4\pi}3a_+^2\\[1pt]
0 & \frac{\alpha_1}{4\pi} (-2\beta_{0,\text{U(1)}}+4a_+^2)&\frac{\alpha_1}{4\pi}a_+^2\\[1pt]
\frac{\alpha_2}{4\pi}4a_+^2&\frac{\alpha_1}{4\pi}4a_+^2&\frac{\alpha_1}{4\pi} (-2\beta_{0,\text{U(1)}}+a_+^2)+\frac{\alpha_2}{4\pi}3a_+^2
\end{pmatrix}, \label{GammaBB}
\end{align}
with 
\begin{align}
\label{GammaAAdiag1}
\gamma_{AA}^{(1)}&=-4\gamma_\text{cusp}\ln\frac{\mu}{2\mDM}+\frac{\alpha_2}{4\pi}(-2\beta_{0,\text{SU(2)}}-8i\pi+4a_+^2)\\
\label{GammaAAdiag2}
\gamma_{AA}^{(2)}&=-4\gamma_\text{cusp}\ln\frac{\mu}{2\mDM}+\frac{\alpha_2}{4\pi}(-2\beta_{0,\text{SU(2)}}-12+4i\pi +4a_+^2)\\
\label{GammaAAdiag3}
\gamma_{AA}^{(3)}&=-4\gamma_\text{cusp}\ln\frac{\mu}{2\mDM}+\frac{\alpha_2}{4\pi}(-2\beta_{0,\text{SU(2)}}-8i\pi)+\frac{\alpha_1}{4\pi}4a_+^2\\
\label{GammaAAdiag4}
\gamma_{AA}^{(4)}&=-4\gamma_\text{cusp}\ln\frac{\mu}{2\mDM}+\frac{\alpha_2}{4\pi}(-2\beta_{0,\text{SU(2)}}-8i\pi+3a_+^2)+\frac{\alpha_1}{4\pi}a_+^2,\\
\label{GammaABdiag1}
\gamma_{AB}^{(1)}&=-2\gamma_\text{cusp}\ln\frac{\mu}{2\mDM}+\frac{\alpha_2}{4\pi}(-\beta_{0,\text{SU(2)}}-4+3a_+^2)+\frac{\alpha_1}{4\pi}(-\beta_{0,\text{U(1)}}+a_+^2)\\
\label{GammaABdiag2}
\gamma_{AB}^{(2)}&=-2\gamma_\text{cusp}\ln\frac{\mu}{2\mDM}+\frac{\alpha_2}{4\pi}(-\beta_{0,\text{SU(2)}}-4+2a_+^2)+\frac{\alpha_1}{4\pi}(-\beta_{0,\text{U(1)}}+2a_+^2),
\end{align}
where $\mu$ (not to be confused with the  Higgsino mass parameter) denotes the renormalization scale associated with dimensional regularization. The cusp piece is
\begin{align}
\gamma_\text{cusp}&=\frac{\alpha_2}{4\pi}\gamma_{\text{cusp}}^{(0)}+\left(\frac{\alpha_2}{4\pi}\right)^2\gamma_{\text{cusp}}^{(1)}\,,
\end{align}
with
\begin{align}
&\hspace*{1cm} \gamma_{\text{cusp}}^{(0)}=4\,,\\
&\hspace*{1cm} \gamma_{\text{cusp}}^{(1)}=\left(\frac{268}{9}-\frac{4\pi^2}{3}\right)c_2(ad)-\frac{80}{9}n_G-\frac{16}{9}n_H\,,
\end{align}
$c_2(ad)=2$, and $n_G=3$ the number of fermion generations. The parameter $a_+$, which is introduced in appendix \ref{app:EWSB}, is related to the Higgs-Higgsino-gaugino interactions and reads
\begin{align}
\label{eq:aplus}
a_+ = \begin{cases} \;
\displaystyle \frac{s_{\alpha_H} +\mathrm{sign}(\mu)
c_{\alpha_H}}{\sqrt{2}} & \;n_H=1 \\
\;1 & \;n_H = 2.
\end{cases}
\end{align}
Therefore all the terms dependent on $a_+$ are absent in pure models. The one-loop running constants of the SU(2) and U(1) couplings are 
\begin{align}
\beta_{0,\text{SU(2)}}=\frac{22}{3}-\frac{n_H}{6}-\frac{4n_G}{3},\quad \beta_{0,\text{U(1)}}=-\frac{n_H}{6}-\frac{20n_G}{9}\,.
\end{align}

The renormalization group evolution of the gauge-couplings $\alpha_{1,2}(\mu)$ is computed using their two-loop RGEs, 
since these couplings appear together with the 
one-loop anomalous dimensions. The QCD and Yukawa 
couplings enter only implicitly 
in the two-loop terms of these RGEs and can themselves by evolved 
with one-loop accuracy.  The Higgs self-coupling does not 
contribute within this accuracy. 
To account for the possibility that the 
second Higgs doublet has mass below the scale of the 
neutralino masses $m_\chi$, we evolve from $m_Z$ to $M_A$ in the 
SM and from $M_A$ to $2 m_\chi$ in the two-Higgs doublet model \cite{BhupalDev:2014bir}. The initial values of the couplings at the electroweak scale  $m_Z =
91.1876\,\text{GeV}$ are taken  from~\cite{Beneke:2019vhz}: $\hat \alpha_2(m_Z) = 0.0350009$, $\hat \alpha_3(m_Z) = 0.1181$, $\hat s^2_W (m_Z) = \hat{g}^2_1 /(\hat{g}^2_1
+ \hat{g}^2_2 )(m_Z) = 0.222958$, $\hat\lambda_t(m_Z) = 0.952957$  in the $\overline{\text{MS}}$ scheme. In addition, we include the bottom
Yukawa coupling $ \hat\lambda_b(m_Z)=0.016447$ computed, analogous to the top-quark coupling, via tree-level relations from
$\overline{m}_b(m_Z)=2.819234\,\text{GeV}$.


\subsection{Extension to non-degenerate models}
\label{sec:nondegenerate}

Strictly speaking, the theory developed in section~\ref{sec:EFT} assumes that the gauginos and Higgsinos are all degenerate within $\delta m_X\,\lesssim\, m_W$. This means that in a rigorous EFT expansion all masses $M_1,\,M_2,\,|\mu|$ are to be expanded around $ \mDM$. However, this causes large errors in the tree-level annihilation matrices $\Gamma_{IJ}^\text{EFT}$ when mass splittings are non-negligible. This work aims to cover arbitrary masses of the bino, wino and Higgsino in the TeV range, including cases when some of the two-particle states are essentially decoupled, hence the above approximation must be improved.
We can actually introduce the full kinematic mass dependence at LO in the chargino channels, making use of the tree-level annihilation matrices with full kinematics, $\Gamma_{IJ}^{\rm exact}$, which were computed in \cite{Beneke:2012tg}, by simply rescaling the NLL result by the ratio  $\Gamma_{IJ}^{\rm exact} / \Gamma_{IJ}^\text{EFT}$. 
For the neutralino annihilation channels $IJ$ this is not possible as there is no tree-level annihilation into photons. Instead, we correct here for the leading dependence of the kinematic factor in \eqref{GammaSoft}, which stems from the approximation $1/M_{IJ}^2 \approx 1/(4\mDM^2)+\mathcal{O}(m_W/\mDM^3)$, where $M_{IJ}=M_{(i_1,i_2)(j_1,j_2)}=\frac{1}{2}(m_{i_1}+m_{i_2}+m_{j_1}+m_{j_2})$. The improved annihilation matrix is
\begin{align}\label{def:GammaImp}
\Gamma^{\NLL,\text{imp}}_{IJ}(\mu_h,\mu_j,\mu_s,\nu_h,\nu_s,E^\gamma_\text{res})\equiv\mathcal{G}_{IJ}\,\Gamma^\NLL_{IJ}(\mu_h,\mu_j,\mu_s,\nu_h,\nu_s,E^\gamma_\text{res}),
\end{align}
where the improvement factor $\mathcal{G}$ to correct for the mass dependence is given by 
\begin{align}
\mathcal{G}_{IJ}=\begin{cases}
4\mDM^2/M_{IJ}^2 & \text{if } \Gamma_{IJ}^\text{EFT}(\mu_h,\nu_h)=0,\\[0.2cm]
\frac{\Gamma_{IJ}^{\rm exact}}{\Gamma_{IJ}^\text{EFT}(\mu_h,\nu_h)} & \text{else},
\end{cases}
\label{eq:gijfactor}
\end{align}
and  $\Gamma^{\rm exact}_{IJ} = 2\Gamma_{IJ}^{\chi_{i_1}\chi_{i_2}\to \gamma\gamma \to \chi_{j_2}\chi_{j_1}}(^1S_0)+\Gamma_{IJ}^{\chi_{i_1}\chi_{i_2}\to \gamma Z \to \chi_{j_2}\chi_{j_1}}(^1S_0)$ is defined as a sum of exclusive annihilation channels from \cite{Beneke:2012tg}, with exact kinematics of the indicated process.

\begin{figure}[t]
\begin{center}
\includegraphics[width=0.65\textwidth]{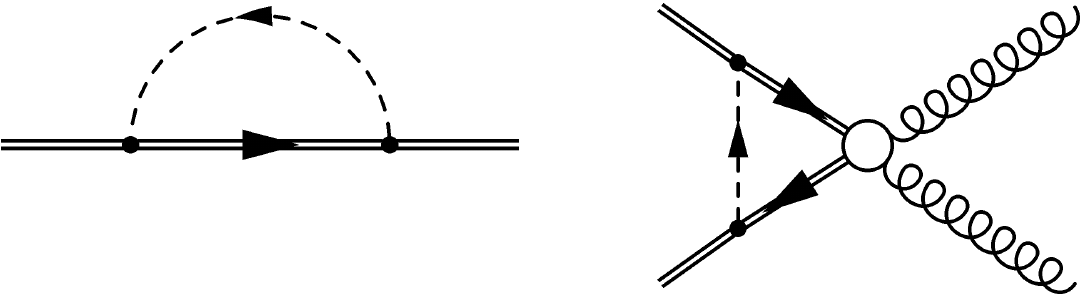}
\caption{Example diagrams for Higgs-exchange contributions to the one-loop anomalous dimension.}
\label{fig:1loop}
\end{center}
\end{figure}

The neutralino mass splittings further enter in the resummation of logarithms from one-loop diagrams involving two different bino, wino or Higgsino gauge multiplets, i.e. diagrams involving the Higgs-Higgsino-gaugino interactions, as depicted in figure \ref{fig:1loop}. 
Generally, in Higgs-loop processes the logarithmic enhancement is cut-off in the infrared by the mass of the exchanged Higgs boson or the mass difference of the Higgsino and gaugino, whichever is larger. In the presence of a large mass 
difference $\delta m_X \gg m_W$,
the denominator of the internal fermion propagator is dominated by $\delta m_X$ in the infrared 
loop momentum region of the full-theory diagram, 
such that the large logarithm $\ln\frac{m_\chi}{m_W}$ disappears. In the EFT calculation of the ultraviolet 
anomalous dimension in dimensional regularization, however, one sets the infrared 
scale $\delta m_X$ to zero, resulting in a contribution to the anomalous dimension 
even when $\delta m_X$ is large. This is a consequence of the well-known feature of non-decoupling in dimensional regularization, which requires that we manually switch off resummation in these terms of the anomalous dimension  by a decoupling function when  $\delta m_X$ becomes large. More precisely, we switch off the anomalous dimension when the internal mass is larger than the external one. In the opposite case, the contribution is automatically suppressed by the small mixing of the heavy external neutralino gauge eigenstate with the lightest physical neutralino mass eigenstate as well as the factor 
$\mathcal{G}_{IJ}$ from \eqref{eq:gijfactor}. We describe the details of the implementation in appendix~\ref{app:Matching}.

The approximation $M_1,\, M_2,\, |\mu| \approx \mDM$ in the arguments of the cusp logarithms of the hard anomalous dimension \eqref{def:Gamma} is unproblematic within the targeted precision even in the case of very different mass parameters, which can be seen from the relative difference of the Wilson coefficients expanded to leading order in 
$\Delta m$, the difference between the sum of the masses of the 
heavy fermion fields in the operator and $2\mDM$,
\begin{align}
\frac{C_i^\NLL}{C_i^\NLL\big|_{\ln 2 \mDM \to \,\ln (2\mDM+\Delta m)}} - 1  
= n_i^{SU(2)} \frac{\gamma_{\text{cusp}}^{(0)}}{2\beta_{0,\text{SU(2)}}}\frac{\Delta m}{\mDM}\ln\frac{\alpha_2(\mu_s)}{\alpha_2(\mu_h)} + \mathcal{O}\left( \alpha_2, \frac{\Delta m^2}{m_\chi^2}\right).
\end{align}
Here, $n_i^{SU(2)}$ is the number of (anti-)collinear SU(2) gauge bosons of the respective operator $\mathcal{O}_i$. 
This difference is power-suppressed in $\Delta m/m_\chi$ and thus negligible for mass splittings $\delta m_X \;\lessim \;m_W$. For larger mass splittings, the incorrect cusp argument of the logarithm still has a negligible effect, since the mixing into the other operator is inhibited by the procedure discussed above while the direct contribution of the operator with the heavy neutralino gauge eigenstates on the lighter physical neutralino mass eigenstate is also suppressed for the same reasons as above. In total this approximation amounts to a difference of 1-2\% for DM masses in the TeV range.


\section{Numerical analysis of resummation}
\label{sec:results}

The resummation described in the previous chapters has been 
implemented in a custom \texttt{Mathematica} code. It uses 
pre-existing work \cite{Beneke:2016ync, Beneke:2016jpw} to include 
the Sommerfeld effect in our computations and to analyze the DM 
relic density and total DM annihilation cross section. This allows us to compute the semi-inclusive 
annihilation cross section in almost the entire parameter range of 
neutralino DM models, excluding only DM annihilation through 
a resonance.\footnote{See \cite{Beneke:2022rjv} for a discussion 
of Sommerfeld enhancement of resonant annihilation.}

\subsection{Neutralino DM parameter space and benchmark models}

To highlight the effect of resumming large logarithms in ${\chi^0\chi^0\rightarrow\gamma+X}$ annihilation up to next-to-leading logarithmic order, we selected a set of MSSM parameter-space points. Table~\ref{tab:DataPh} summarizes these benchmark models that cover viable mixed TeV-scale neutralino cases. The naming of the model includes the letter
B (Bino), H (Higgsino) or W (wino), whenever the lightest-neutralino admixture of the respective gauge eigenstate is larger than 5\%. Since the novelty of this work lies in the simultaneous treatment of bino, wino and Higgsinos we focus on models with mixed LSP content. 
We note that out of the 23 studied models only four are fully-mixed (BHW) with this definition, since direct detection constraints already exclude relatively small Higgsino admixtures (but not the pure Higgsino). The full specification of the models in terms of their MSSM parameters is given in appendix~\ref{app:data}. 

\begin{table}[p]\begin{center}
\begin{tabular}{r|cc|cc|c}

model & $m_{LSP}$ & $m_{\chi^\pm_1}$  & $\left<\sigma v_{rel}\right>^{SE+NLL}_{\gamma X}$   & $\frac{\left<\sigma v_{rel}\right>^{SE+NLL}_{\gamma X} }{ \left<\sigma v_{rel}\right>^{SE}_{\gamma X}} $ & $\frac{\left<\sigma v_{rel}\right>_{tot}^{SE}}{\left<\sigma v_{rel}\right>_{tot}^{LO}}$  \\
&{\small [$GeV$]}&{\small{[$GeV$]}}&{\small[$cm^3s^{-1}$]}&&\\
\hline
\emph{pure models}\,\,\,\quad\quad &&&&&\\
\hline\hline
H\, & 1111.6 & 1112.4 & $1.50\cdot 10^{-28}$ & $\bf{ 0.82 }$& 1.58 \\

W\, & 2849.5 & 2849.6 & $8.04\cdot 10^{-26}$ & $\bf{ 0.60 }$& 83.6 \\[3pt]
\emph{doubly mixed}\quad\quad &&&&& \\

\hline\hline

BH\, & 1065.4 & 1069.8 & $7.58\cdot 10^{-29}$ & $\bf{ 1.16 }$& 1.27 \\

BW\, & 2141.7 & 2143.8 & $1.46\cdot 10^{-27}$ & $\bf{ 0.64 }$& 5.28 \\

BW-e\, & 1825.7 & 1830.5 & $1.76\cdot 10^{-28}$ & $\bf{ 0.67 }$& 2.25 \\

BW-2520\, & 2516.4 & 2516.9 & $1.57\cdot 10^{-25}$ & $\bf{ 0.61 }$& 168 \\

BW-e-nh2\, & 2054.0 & 2056.1 & $1.80\cdot 10^{-28}$ & $\bf{ 0.64 }$& 3.57 \\

HW\, & 2830.9 & 2831.1 & $2.60\cdot 10^{-25}$ & $\bf{ 0.58 }$& 322 \\

HW-nh2\, & 2912.2 & 2912.4 & $1.42\cdot 10^{-25}$ & $\bf{ 0.55 }$& 192 \\[3pt]

\emph{fully mixed}\quad\quad\quad &&&&& \\

\hline\hline

BHW-mix\, & 1916.1 & 1922.0 & $1.14\cdot 10^{-28}$ & $\bf{ 0.64 }$& 2.11 \\

BHW-mix2 & 1966.0 & 1971.3 & $1.66\cdot 10^{-28}$ & $\bf{ 0.64 }$& 2.34 \\

BHW-mass\, & 1621.1 & 1632.4 & $1.66\cdot 10^{-29}$ & $\bf{ 0.70 }$& 1.39 \\

BHW-nh2\, & 1797.1 & 1808.5 & $3.02\cdot 10^{-29}$ & $\bf{ 0.66 }$& 1.64 \\[3pt]

\emph{additional }\quad\quad\quad &&&&& \\

\hline\hline

$^\ast$B & 2144.9 & 6997.5 & $4.36\cdot 10^{-36}$ & $\bf{ 2.21\cdot 10^{5} }$& 1.01 \\

$^\ast$BW-coan\, & 2144.9 & 2147.6 & $2.63\cdot 10^{-31}$ & $\bf{ 0.64 }$& 1.17 \\

$^\ast$H+\, & 1111.4 & 1112.2 & $1.53\cdot 10^{-28}$ & $\bf{ 0.81 }$& 1.58 \\

H2\, & 1236.6 & 1238.9 & $1.05\cdot 10^{-28}$ & $\bf{ 0.78 }$& 1.50 \\

BH-undet\, & 1296.1 & 1316.0 & $3.31\cdot 10^{-30}$ & $\bf{ 0.86 }$& 1.15 \\

BW-nfw\, & 2073.6 & 2075.7 & $3.70\cdot 10^{-28}$ & $\bf{ 0.64 }$& 4.07 \\

BW-ce\, & 2284.8 & 2285.7 & $7.14\cdot 10^{-27}$ & $\bf{ 0.63 }$& 15.2 \\

BW-2670\, & 2663.1 & 2664.0 & $3.77\cdot 10^{-26}$ & $\bf{ 0.58 }$& 54.8 \\

BW-nh2\, & 2436.0 & 2436.7 & $3.29\cdot 10^{-26}$ & $\bf{ 0.62 }$& 41.3 \\

BW-ce-nh2\, & 2162.9 & 2164.1 & $3.49\cdot 10^{-27}$ & $\bf{ 0.63 }$& 9.43 \\
\end{tabular}
\caption{\label{tab:DataPh}
Summary of results for the 23 selected MSSM benchmark models. See explanation in 
the main text.}
\end{center}\end{table}

All benchmark points apart from ``B", ``H+" and ``BW-coan" (marked by an asterisk) yield DM relic densities between 
0.1182 and 0.1191 in agreement with the observed value.  We used \texttt{micrOMEGAs} \cite{Belanger:2020gnr} (version 5.2.7.a) to ensure that all models evade current bounds from direct detection and collider experiments and thus, apart from the three mentioned exceptions, represent viable DM models.\footnote{Indirect detection constraints are not applied as we show 
the annihilation cross section into $\gamma+X$ in figure~\ref{fig:SE+NLLcs} below.}
Further, we cross-checked our results for the relic density and exclusive annihilation cross sections, with the Sommerfeld effect and resummation turned off, with \texttt{micrOMEGAs} and find agreement within $2\%$ as long as the $t\bar{t}$ final state channel is negligible. Where it is significant, the $t\bar{t}$ channel causes up to $\mathcal{O}(10\%)$ differences in the total cross section and relic density. The difference arises because the running of the top-quark mass is neglected in the implementation of the tree-level annihilation matrices $\Gamma^{\rm exact}_{IJ}$ from 
\cite{Beneke:2016ync, Beneke:2016jpw}. The $t\bar{t}$ channel is however negligible in most benchmarks, except for mixed bino-Higgsino DM with negligible wino admixture. 
The basic implementation of the Sudakov resummation was validated for the almost pure wino ``W" and Higgsino ``H"  models by comparison with existing codes \cite{Beneke:2018ssm,Beneke:2019vhz}. Agreement to $0.1\%$ was found for these non-degenerate models. Apart from correct numerical and analytical implementations, this validates the decoupling of heavy particles described in the previous section (see appendix \ref{app:Matching}). 

The masses of the heavy sfermions are set to a common large value $m_{sf}\gg \mDM$. The effect of sfermion masses on the relic density has been investigated in \cite{Beneke:2016ync}, and can be noticeable, but we do not expect it to affect the relative effect of Sudakov resummation.
Apart from $m_{sf}$, the model has five further parameters: the gaugino masses $M_{1,2}>0$, the Higgsino mass parameter $\mu$ (which may be negative), the ratio of the Higgs vacuum expectation values (vevs) $\tan\beta$, and the neutral pseudo-scalar Higgs mass $M_{A}$. 

Columns 2 and 3 of table \ref{tab:DataPh} provide the lightest supersymmetric 
particle (LSP) mass  $m_{\chi}$ together with the lightest chargino mass 
$m_{\chi_1^\pm}$. The main results of this work is found in columns 4 and 5. 
The former gives $\left< \sigma v_{rel} \right>_{\gamma X}^{SE+NLL}$ as defined 
in \eqref{NLLann} with photon energy resolution $\Eres=m_W=80.385$~GeV. 
The fifth column (in bold) quantifies the importance of resummation by taking 
the ratio of the semi-inclusive DM annihilation cross section into photons including Sommerfeld enhancement (SE)  and Sudakov resummation at NLL accuracy, $\left< \sigma v_{rel} \right>_{\gamma X}^{SE+NLL}$, over the previous state of the art, $\left< \sigma v_{rel} \right>_{\gamma X}^{SE}$, which included only the former.  For comparison, the last column provides information on the size of the Sommerfeld effect on the total annihilation cross section by 
taking the ratio of $\left<\sigma v_{rel}\right>_{tot}^{SE}$ 
to the corresponding Born cross section $\left<\sigma v_{rel}\right>_{tot}^{LO}$. 

\begin{figure}[t]
\begin{center}
\includegraphics[width=.9\textwidth]{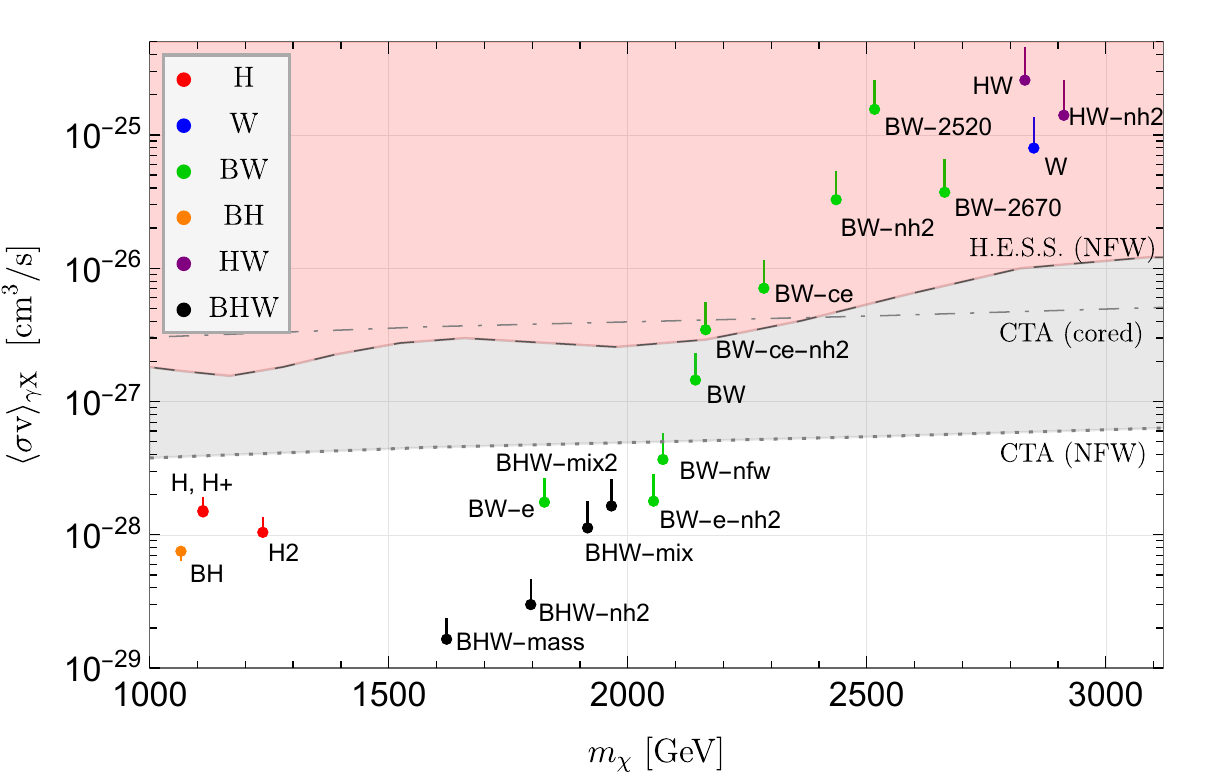}
\caption{$\left<\sigma v_{rel}\right>^{SE+NLL}_{\gamma X}$ (dots) and resummation effect (attached vertical lines) plotted against the LSP mass $m_\chi$ for the models listed in appendix \ref{app:data} and table \ref{tab:DataPh}. The colour-coding indicates the bino ``B", Higgsino ``H", wino ``W" LSP content larger than 5\%. The shaded areas show exclusion limits from H.E.S.S. (dashed) \cite{HESS:2018cbt} and expected exclusions for the CTA experiment (bold, from \cite{Hryczuk:2019nql}), both assuming a NFW profile
for the DM density distribution. The limit weakens by about an order 
of magnitude for the cored Einasto profile.
\label{fig:SE+NLLcs}
}
\end{center}
\end{figure}

\subsection{Discussion}
\label{sec:resultsNLL}

Figure \ref{fig:SE+NLLcs} shows the present-day semi-inclusive annihilation cross section $\left\langle\sigma v_{rel} \right\rangle_{\gamma X}$ \eqref{NLLann} 
at $v_{rel}=0.001$ against the mass of the DM particle for the 23 models from table~\ref{tab:DataPh}. The dot represents the value of the resummed value, $\left<\sigma v_{rel}\right>^{SE+NLL}_{\gamma X}$, which is our main result, while the reduction effect of resummation is indicated by the attached vertical line, the upper end of which is  $\left<\sigma v_{rel}\right>^{SE}_{\gamma X}$ without Sudakov resummation.

Between 1.5\,TeV and 3\,TeV, the correct relic abundance requires that cross sections increase with $\mDM$ such that DM is sufficiently depleted before freeze-out. Consequently, the benchmarks sort themselves roughly by their wino admixture with mainly-bino fully mixed models (black) at smaller and pure wino-like models (blue) at larger masses. The LSP mixing is caused by the off-diagonal mixing entries in the neutralino mass matrix (\ref{MNNR}). Since wino and bino do not mix directly but only through their interactions with the Higgsino, one can steer bino-wino models (green) from a truly mixed LSP to an almost pure LSP with co-annihilations of a degenerate gaugino simply by increasing $|\mu|$, effectively decoupling the Higgsino. 
Note that this is possible because the decoupling at large mass splittings introduced in appendix \ref{app:Matching} does not affect the diagonalization of the mass matrices. 

From column 5 (bold) of table \ref{tab:DataPh} we infer that with few exceptions, 
electroweak Sudakov 
resummation reduces the annihilation into photons by  $20\%$ for the pure Higgsino 
model at $\mDM\approx 1.1\,$TeV up to $45\%$ for wino-dominated models around 
2.8~TeV. The resummation effect generally grows slightly with the DM mass due 
to the larger ratio to the electroweak scale, and consequently larger logarithms. 
A notable exception is the bino-Higgsino mixed model "BH" (orange) for which
resummation yields a $+11\%$  enhancement. A similar effect was found for the 
pure Higgsino at smaller mass $\mDM\lesssim 1\,$TeV \cite{Beneke:2019gtg}, 
which is shifted here to larger mass due to the bino admixture.\footnote{
The abnormally large resummation effect for the bino-dominated model ``B'' is 
related to the very small tree-level annihilation cross section due to an 
electromagnetic cancellation, and the absence of a Sommerfeld effect for 
binos. The cancellation is lifted already in the leading-logarithmic 
resummation of radiative corrections, which by far exceed the tree-level 
contributions even for small admixtures in these heavy bino-dominated models.}

The 23 models include five cases (suffix -nh2) that feature two light Higgs 
doublets. The effect of $M_A \;\lessim\; 2 \mDM$ is primarily to open 
additional annihilation channels into $H^0$, $A^0$ and $H^\pm$, increasing 
the cross section. Consequently, for such light Higgs models to produce the correct 
relic density, they are generically located at slightly larger masses, compensating 
for the additional DM depletion during freeze-out. The second Higgs doublet enters  
the anomalous dimension through the quantity $a_+$ defined in \eqref{eq:aplus} and 
the evolution of the gauge couplings.

Figure \ref{fig:SE+NLLcs} shows (shaded) the regions 
excluded by null-observations of a $\gamma$ signal from 
H.E.S.S. (dashed) \cite{HESS:2018cbt} as well
as simulated expected exclusions  \cite{Hryczuk:2019nql} 
from the upcoming CTA experiment (dotted), both assuming a 
NFW profile for the DM density distribution.\footnote{
We used the simulated data for the CTA sensitivity to 
monochromatic gamma ray lines provided in the ancillary files to 
\cite{Hryczuk:2019nql} and extracted the experimental 
constraints from H.E.S.S. obtained after 10 years of 
observation from \cite{HESS:2018cbt}. These bounds on the 
monochromatic gamma ray line have been multiplied by a 
factor 2 to account for plotting the semi-inclusive process 
$\langle \sigma v\rangle_{\gamma X}$ on the vertical axis.} These 
apparently exclude the high-mass benchmark models, but  
it it must be kept in mind that the cross-section limit 
weakens by up to an order of magnitude in case of a cored 
DM density profile (as shown by the expected 
cored Einasto-profile CTA exclusion). 

Corrections to the NLL results presented here arise from power-suppressed 
terms in $m_W/m_\chi$ and higher-order loop corrections to the resummation 
functions. The former have been investigated for the pure-Higgsino case 
\cite{Beneke:2019gtg} and were found to be below the percent-level even 
for low masses $m_\chi\sim 1$~TeV. For the pure wino and Higgsino, NLL' 
resummation is available \cite{Beneke:2019vhz,Beneke:2019gtg}, which differs 
from NLL by adding the non-logarithmic one-loop corrections to all resummation 
functions. For the full MSSM, computing all hard functions to the one-loop 
order would be rather involved. Comparing NLL' and NLL accuracy 
for the pure models, one concludes that the scale uncertainty is significantly 
reduced at NLL', but for the scale choices adopted in the present work, 
the central value is almost the same in the 1-2~TeV mass region. The NLL result 
is larger than the NLL' one by a few percent for larger masses. We therefore 
estimate that the uncertainty on the resummation correction is a few percent 
with a tendency of radiative corrections to strengthen the resummation effect.

\section{Summary}
\label{sec:conclusions}

In this paper, we presented for the first time the neutralino annihilation cross section into photons including NLL resummation of electroweak 
Sudakov logarithms for arbitrarily mixed TeV-scale neutralino DM. 
For a series of benchmark models consistent with the observed relic 
abundance and direct detection constraints, as well as other constraints on the MSSM parameter space, Sudakov resummation gives rise to a reduction of the annihilation cross section, which ranges from about $20\%$ for Higgsino-dominated DM at smaller masses around 1~TeV to $45\%$ DM with larger wino admixture at heavier masses around $3$ TeV. 

As a rough rule of thumb, the effect of NLL resummation of Sudakov logarithms 
on the semi-inclusive annihilation cross section can be captured 
by multiplying the annihilation rate without resummation by a factor 0.6 to 0.65 
within the few TeV mass range under consideration. This makes Sudakov resummation 
an essential part of any precise annihilation cross section computation, 
which should be taken into account in translating cross section limits from 
the non-observation of the line-like signal from DM annihilation into the 
MSSM parameter space.
%

\subsubsection*{Acknowledgements}
We thank Kai Urban, Caspar Hasner and Martin Vollmann as well as Andrzej Hryczuk and Krzysztof Jod{\l}owski for helpful discussions. The support of Stefan Recksiegel and Kai Urban in implementing and checking the numerical code is greatly appreciated. MB thanks the Albert Einstein Center at the University of Bern for hospitality when this work was completed. 
This work has been supported by the Collaborative Research Center ``Neutrinos and Dark Matter in Astro- and Particle Physics'' (SFB 1258). CP is supported by the Spanish Ministry grant PID2019-106080GB-C21.



\appendix

\section{The Higgs sector in the NREFT and EWSB}
\label{app:EWSB}

In this appendix we discuss the different paths of how to introduce the electroweak symmetry breaking:
\begin{align*}
\textbf{Path 1: } \quad& \text{MSSM}\quad \Rightarrow\quad \text{NR} \text{ MSSM}\quad\Rightarrow\quad \text{NR-}\slashed{\text{EW}} \text{ MSSM}\\
\textbf{Path 2: } \quad& \text{MSSM}\quad \Rightarrow\quad \slashed{\text{EW}} \text{ MSSM}\quad\Rightarrow\quad \text{NR-}\slashed{\text{EW}} \text{ MSSM}
\end{align*} 
As discussed in section~\ref{sec:EFT}, the transition to the non-relativistic EFT as well as the decoupling of a heavy Higgs doublet (if present) naturally happen before EWSB (path 1). However, one could also reach the non-relativistic theory with broken electroweak symmetry ($\slashed{\text{EW}}$) by first going through EWSB and then taking the NR limit (path 2). 
Both paths are equivalent as long as power expansions in the EFT construction are done consistently to the same order. This appendix provides details on the decoupling procedure and compares path 1 to the more common approach of path 2. 

\subsection{Path 1: NR limit before EWSB}

Before EWSB the NR limit for gauginos and Higgsinos is simple as there is no mixing between different gauge multiplets. The only difficulty arises from the fact that the Higgsino mass parameter $\mu$ of the MSSM can have either sign, while the mass parameter in the kinetic term of a NREFT must always be positive. This problem can be addressed by defining the four-component Dirac Higgsino as $\Psi_H \equiv \left(\mathrm{sign}(\mu)\,\psi_1,(-\epsilon\psi_2^c)\right)$ 
in terms of the Weyl fermions $\psi_{1,2}$. This yields the usual Dirac fermion Lagrangian $\overline{\Psi}_H(i\slashed D - |\mu|)\Psi_H$, which can be matched to a non-relativistic Lagrangian at the price  of an explicit dependence on sign$(\mu)$ in the Higgs-Higgsino-gaugino interactions \eqref{LagFull}.

Starting from the MSSM Higgs sector, using the notation and conventions of \cite{Rosiek:1995kg}, we diagonalize the mass matrix of the SU(2) Higgs doublets $(H_1,H_2)$ before EWSB by means of the unitary rotation
\begin{align}\label{hH}
\begin{pmatrix}
h\\H
\end{pmatrix}=\begin{pmatrix}
-c_{\alpha_H}& s_{\alpha_H}\\
\phantom{-}s_{\alpha_H}& c_{\alpha_H}
\end{pmatrix}\begin{pmatrix} \epsilon H_1^*\\H_2
\end{pmatrix}
\end{align}
to the doublets $(h,H)$, 
with the angle $\alpha_H$ defined in terms of the MSSM parameters as
\begin{align}\label{def:alphaH}
t_{\alpha_H} &= \frac{-2m_{12}^2}{m_{H_2}^2-m_{H_1}^2+\sqrt{4m_{12}^2+(m_{H_2}^2-m_{H_1}^2)^2}}\,.
\end{align}
The mass and quartic terms in the Higgs Lagrangian are now given by
\begin{align}\label{LmBEWSB}
\mathcal{L}_{\rm Higgs}=&-\left(M_h^2 h^\dagger h+M_H^2H^\dagger H\right)-\frac{g_2^2}{2}\left((h^\dagger h)(H^\dagger H)-(h^\dagger H)(H^\dagger h)\right)\nn\\
&-\frac{g_1^2+g_2^2}{8}\left(c_{2\alpha_H} (h^\dagger h-H^\dagger H)-s_{2\alpha_H}(h^\dagger H+H^\dagger h)\right)^2
\end{align}
with 
\begin{align}
M_{h/H}^2&\equiv\frac{1}{2}(M_A^2\pm \frac{2m_{12}^2}{s_{2\alpha_H}}),
\quad\text{ where }
\quad  M_A^2\equiv m_{H_1}^2+m_{H_2}^2+2|\mu|^2\,.
\end{align}
$M_A$ is introduced here as a convenient abbreviation of parameters, 
however after EWSB it will correspond to the pseudo-scalar Higgs mass. If $M_A\gg m_Z$, after decoupling the heavy $H$ field, the remaining $h$ coincides with  the SM Higgs doublet, and 
$\frac{-2m_{12}^2}{s_{2\alpha_H}} \approx M_A^2$ in order that $M_h \sim m_Z$. 
Hence the decoupling limit corresponds to
\begin{align}\label{dec-cond}
M_A^2\gg m_Z^2\quad \text{ and }\quad 
s_{2\alpha_H}\to  -\frac{2m_{12}^2}{M_A^2}.
\end{align}

Turning to the construction of the NR MSSM with unbroken electroweak symmetry, we first perform the matching of the MSSM Higgs sector onto the NREFT. The relevant MSSM gaugino-Higgsino-Higgs interaction terms read
\begin{align}\label{LagFull}
\mathcal{L}_{\rm MSSM} \supset
&-\sqrt{2}H_1^\dagger\left(g_2\overline{\chi}_A^aT^a-\frac{g_1}{2}\overline{\chi}_B\right)P_L\,\mathrm{sign}(\mu)\chi_H\nonumber\\
&+\sqrt{2}H_2^T\epsilon\left(g_2\overline{\chi}_A^aT^a-\frac{g_1}{2}\overline{\chi}_B\right)P_R\,\chi_H+ \mbox{ h.c.}\,,
\end{align}
where the wino $\chi_A$ and bino $\chi_B$ Majorana spinors form a triplet ($\mathbf{3}_0$) and singlet ($\mathbf{1}_0$) of SU(2)$_L\times$U(1)$_Y$, respectively. The charged Dirac field of the Higgsino $\chi_H$ transforms under (${\bf 2}_{-1/2}$) and includes both superpartners of the scalar Higgs doublets, which carry opposite hypercharges (${\bf 2}_{-1/2}$ for $H_1$ and ${\bf 2}_{1/2}$  for $H_2$). The matrices $P_{R/L}=(1\pm\gamma_5)/2$ are the right-/left-chiral projectors defined in the Weyl basis. When projecting onto non-relativistic fermion field bilinears, the parity-odd combination $P_L-P_R$ is suppressed in the non-relativistic expansion with respect to the even one, $P_L+P_R$. It is thus convenient to decompose the interactions in the NREFT Lagrangian using these parity projectors as well as linear combinations of the Higgs fields with a definite parity eigenvalue. We therefore define 
\begin{align}\label{def:Hpm2}
H_\pm &=
\frac{s_{\alpha_H}\pm\mathrm{sign}(\mu) c_{\alpha_H}}{\sqrt{2}} \, h +
(n_H-1)\frac{c_{\alpha_H}\mp\mathrm{sign}(\mu) s_{\alpha_H}}{\sqrt{2}} \, H 
\,,
\end{align}
where now $ H_+$ is parity-even and couples to Higgsinos and gauginos as a scalar, while $H_-$ is parity-odd and couples as a pseudo-scalar. When the additional MSSM Higgs states have masses closer to the electroweak scale than $m_\chi$, they must be included as dynamical fields in the NR Lagrangian. Otherwise they must be integrated out. We have implemented the tree-level decoupling of the Higgs doublet $H$ when it is heavy via the factor $(n_H-1)$ in \eqref{def:Hpm2}, where $n_H$ denotes the number of light Higgs doublets. Note that this decoupling is a separate effect from the suppression of the NR coupling of $H_-$ to gauginos and Higgsinos. Dropping the terms which will be suppressed  after matching to the NREFT amounts to replacing $P_{L,R}\to \frac{1}{2}$ and (\ref{LagFull}) simplifies to
\begin{equation}
\mathcal{L}_{\rm MSSM} \supset
(H_+^T\epsilon) \left(g_2\overline{\chi}_A^aT^a-\frac{g_1}{2}\overline{\chi}_B \right)\chi_H + \mbox{h.c.}\,.
\end{equation}
In the presentation of the anomalous dimensions in the main text, the dependence 
on $n_H$ was compactly included by defining the parameter 
\begin{equation}
a_+ = \begin{cases}
\displaystyle \;\frac{s_{\alpha_H}+\mathrm{sign}(\mu)c_{\alpha_H}}{\sqrt{2}} & \text{if}~n_H=1, \\
\;1 & \text{if}~n_H=2.
\end{cases}
\end{equation}
For $n_H=1$, this introduces a dependence on $\alpha_H$ in computations involving 
Higgs-Higgsino-gaugino couplings, which arises from the 
coupling to the left-over Higgs field $h$ in \eqref{def:Hpm2}.

We now consider EWSB in the NR MSSM. The electromagnetically neutral 
Higgs field components acquire a vacuum expectation value (vev). For 
$H_+$, 
\begin{align}
\langle H_+\rangle
& = \frac{s_{\alpha_H}+\mathrm{sign}(\mu) c_{\alpha_H}}{\sqrt{2}} \langle h \rangle - (n_H-1) \frac{c_{\alpha_H}-\mathrm{sign}(\mu) s_{\alpha_H}}{\sqrt{2}}\langle H\rangle
\equiv\frac{1}{\sqrt{2}}\begin{pmatrix}
0\\ v_+ 
\end{pmatrix},
\end{align}
where
\begin{equation}
v_+ = \frac{v_2+\mathrm{sign}(\mu)v_1}{\sqrt{2}}
\end{equation}
in terms of the vevs $v_{1,2}$ of the original MSSM fields 
$H_{1,2}$. 
The vev generates non-diagonal mass terms for the gauginos and Higgsinos in \eqref{LagNRg}. Defining $\xi_A^\pm\equiv\frac{\xi_A^1\mp i\xi_A^2}{\sqrt{2}}$, we first introduce the matrix $G$ which rotates the NR fields of \eqref{def:chi} from the gauge- to electromagnetic charge eigenstates via 
\begin{align}
\begin{pmatrix}
\chi^0&\chi^+&\chi^-
\end{pmatrix}^T=G\chi
\end{align}
where 
\begin{align}\label{def:chi0pm}
\chi^0=\begin{pmatrix}
\xi_B&\xi_A^3&\zeta_H^1&\eta_H^1
\end{pmatrix}^T,\quad\quad \chi^+=\begin{pmatrix}
\xi_A^+&\zeta_H^2
\end{pmatrix}^T,\quad\quad \chi^-=\begin{pmatrix}
\xi_A^-&\eta_H^2
\end{pmatrix}^T
\end{align}
collect fields of equal electric charge. In practice, $G$ is mostly trivial since all fields but $\xi_A^{1,2}$ are already charge eigenstates. 

Next, we define the matrix $R$ that diagonalizes the mass matrix in each sector of fixed electric charge, 
\begin{align}
\chi_M = R \begin{pmatrix}
\chi^0&\chi^+&\chi^-
\end{pmatrix}^T =  RG\chi\,,\label{def:chiM}
\end{align}
where $\chi_M$ denote the mass eigenstates.
In path 1, $R^{(\text{P1})}$ has block-diagonal form 
\begin{align}
R^{(P1)}=\begin{pmatrix}\label{def:MP1}
{N^{(P1)}}^{-1}&&\\
&{U^{(P1)}}^{-1}&\\
&&{U^{(P1)}}^{-1}
\end{pmatrix}\,,
\end{align}
where $N^{(P1)}$ is the unitary matrix that diagonalizes the four neutralinos, 
and $U^{(P1)}$ the equivalent for charginos,
\begin{align}
\chi^0=N^{(P1)}\,\chi_M^{0},\quad\quad \chi^\pm=U^{(P1)}\,\chi_M^{\pm}.
\end{align}
At this point, the Lagrangian in \eqref{LagNRg} is expressed in terms of the two-component spinor fields in  
$\chi_M$, which have diagonal mass terms and definite electric charge. 

The mass matrix in path 1 before the diagonalization with $R^{(P1)}$ already 
provides relevant information. To this end, we perform a unitary rotation on $\chi^0$ in \eqref{def:chi0pm} that rotates the neutral NR Higgsino fields $\eta_H^1$, $\zeta_H^1$ into new fields $\xi_{H1,H2}$ by $\frac{\pi}{4}$ and the gauginos $\xi_A$, $\xi_B$  into $\xi_{G1,G2}$ by the Weinberg angle $\theta_W$, i.e. we define 
\begin{align}
\xi_{H1} = \frac{1}{\sqrt{2}}(\eta_H^1+\zeta_H^1),\quad\quad\xi_{H2} = \frac{1}{\sqrt{2}i}(\eta_H^1-\zeta_H^1)\\
\xi_{G1}=-s_W\xi_A^3-c_W\xi_B,\quad\quad \xi_{G2}=c_W\xi_A^3-s_W\xi_B,
\end{align}
where we abbreviated the sine (cosine) of the Weinberg angle as $s_W$ ($c_W$). 
In terms of these fields the Lagrangian \eqref{LagNRg} after EWSB yields the 
neutralino mass matrix 
\begin{align}\label{MNNR}
\mathcal{L}_\chi \supset 
-\frac{1}{2} \, \begin{pmatrix}
\xi_{G1} \\ \xi_{G2} \\ \xi_{H1} \\ \xi_{H2}
\end{pmatrix}^{\!\dagger} \cdot  
\begin{pmatrix}
\delta m_2-c_{W}^2\Delta&-c_{W} s_{W} \Delta &0&0\\
-c_{W}s_{W}\Delta &\delta m_2-s_{W}^2\Delta&\frac{e}{2 s_{W}c_{W}}v_+&0\\
0&\frac{e}{2 s_{W}c_{W}}v_+&\delta m_H&0\\
0&0&0&\delta m_H
\end{pmatrix}
\begin{pmatrix}
\xi_{G1} \\ \xi_{G2} \\ \xi_{H1} \\ \xi_{H2}
\end{pmatrix}
\end{align}
with $\Delta=M_2-M_1$. We observe two features: 1) one of the NR Higgsinos ($\xi_{H2}$) does not mix at all with the other fields and 2) for $\Delta\to 0$ there is another neutralino that does not mix with the others ($\xi_{G1}$). The  lightest neutralino, which is our DM candidate, is always a combination of the remaining two fields $\xi_{G2,H1}$.

\subsection{Path 2: NR limit after EWSB}

Implementing EWSB or the decoupling of heavy Higgs bosons (if present) before taking the NR limit  is the more common procedure, and we omit the detailed description here. For the first step from the full MSSM to the $\slashed{\text{EW}}$ MSSM, we refer again to \cite{Rosiek:1995kg}.
The decoupling limit for a heavy Higgs boson is considered e.g.~in \cite{Haber:1995be,Djouadi:2005gj}. The CP-odd Higgs fields are rotated to their mass eigenstates by the angle $\beta$ defined  as $\tan\beta \equiv 
t_\beta=v_2/v_1$ through the ratio of the vevs of $H_1$, $H_2$. The angle $\alpha_H$ from \eqref{hH} is related 
to the angle $\beta$ (using, e.g.,  (8.1.11) and (8.1.19) of \cite{Martin:1997ns}) by 
\begin{align}
t_{2\beta}=t_{2\alpha_H}\left(1+\frac{m_Z^2}{M_A^2}\right).
\end{align}
{}From \eqref{dec-cond}, we derive that in the decoupling limit 
$\alpha_H \to \beta$, rendering the Higgs masses and couplings 
in both paths equal up to higher-order corrections in $m_Z/M_A$. 

We can now move on to the $\slashed{\text{EW}}$ NREFT. The rotation matrix to the mass eigenbasis $R^{(P2)}$ is
\begin{align}
R^{(P2)}=
\begin{pmatrix}
{N^{(P2)}}^{-1}&&\\
&{U^{(P2)}_+}^{-1}&\\
&&{U^{(P2)}_-}^{-1}
\end{pmatrix},
\end{align}
with the 4$\times$4 matrix $N^{(P2)}$
\begin{align}
N^{(P2)}_{1i} &= \frac{(Z_N)_{1i} + (Z_N^\ast)_{1i}}{2}\,, \quad 
N^{(P2)}_{2i} = \frac{(Z_N)_{2i} + (Z_N^\ast)_{2i}}{2}\,, \nonumber\\
N^{(P2)}_{3i} &=\frac{-(Z_N)_{4i} + \mathrm{sign}(\mu)(Z_N^\ast)_{3i}}{2} \,, \quad 
N^{(P2)}_{4i} =(N^{(P2)}_{3i})^*
\end{align}
and the 2$\times$2 matrices $U_\pm^{(P2)}$
\begin{align}
\left(U_-^{(P2)}\right)_{1k} &= \frac{(Z_-)_{1k} + (Z_+^\ast)_{1k}}{2} \,, \quad 
\left(U_-^{(P2)}\right)_{2k} = \frac{\mathrm{sign}(\mu)(Z_-)_{2k} + (Z_+^\ast)_{2k}}{2}\,,\nonumber\\
U_+^{(P2)} &=(U_-^{(P2)})^\ast\,,
\end{align}
where the mixing matrices $Z_N,Z_\pm$ are the mixing matrices for neutralinos and charginos that can be found in \cite{Rosiek:1995kg}.\footnote{
In \cite{Rosiek:1995kg} it is mentioned that the chargino mass eigenvalues may be kept positive by choosing relative phases in $Z_\pm$ but this is not enforced. However, in NREFT the positive mass requirement is of importance and any negative mass eigenvalues for neutralinos or charginos must be removed by appropriate phase rotations.
}
We find that
\begin{align}
R^{(P2)}=R^{(P1)}+\mathcal{O}\!\left(\frac{m_W}{m_\chi}\right),
\end{align}
and for the neutralino and chargino masses  $m_{\chi_i^{0,\pm}}$ 
\begin{equation}
\label{eq:massesP1P2}
m_{\chi_i^{0,\pm}}^{(P2)} = m_{\chi_i^{0,\pm}}^{(P1)} 
+ \mathcal{O}\!\left(\frac{m_W^2}{m_\chi}\right).
\end{equation}

To see this, we first perform an orthogonal transformation 
of the standard neutralino mass matrix \cite{Rosiek:1995kg} 
and subtract $m_\chi$ times the unit matrix
to bring it into the form 
\begin{align}\label{MNP2}
M^{(P2)\,\prime} =  
\begin{pmatrix}
\delta m_2-c_{W}^2\Delta&-c_{W} s_{W} \Delta &0&0\\
-c_{W}s_{W}\Delta &\delta m_2-s_{W}^2\Delta&\,\frac{e}{2 s_{W}c_{W}}v_+&\frac{e}{2 s_{W}c_{W}}v_-\\
0&\frac{e}{2 s_{W}c_{W}}v_+&\delta m_H&0\\
0&\frac{e}{2 s_{W}c_{W}}v_-&0&-2m_\chi-\delta m_H
\end{pmatrix}\,,
\end{align}
where here, as before in path~1, $m_\chi=\mbox{min}(M_1,M_2,|\mu|)$ and 
\begin{equation}
v_-= \frac{v_2-\mathrm{sign}(\mu)v_1}{\sqrt{2}}\,.
\end{equation}
The characteristic polynomial, from which the eigenvalues of 
the mass matrix \eqref{MNNR} of path 1 are determined, can be written 
as $(\delta m_H-\lambda)\,P_1(\lambda)$. The characteristic 
equation of $M^{(P2)\prime}$ is then expressed as 
\begin{equation}
(2 m_\chi+\delta m_H+\lambda)\left[P_1(\lambda)
-\frac{(\delta m_H-\lambda)(\lambda+c_W^2\Delta-\delta M_2)}
{2 m_\chi+\delta m_H+\lambda}\, 
\frac{e^2 v_-^2}{4 s_W^2 c_W^2}\right] \stackrel{!}{=}0\,.
\end{equation}
Since the second term in square brackets is suppressed 
by the large scale $m_\chi$, while $P_1(\lambda)$ is independent 
of $m_\chi$, it follows that the three eigenvalues 
of $M^{(P2)\,\prime}$ corresponding to the zeros of $P_1(\lambda)$ differ 
by at most $\mathcal{O}(m_W^2/m_\chi)$. The fourth 
eigenvalue plus $m_\chi$ is $m_{\chi_i^{0}} = m_\chi+\delta m_H=|\mu|$ 
for P1 and close to $-|\mu|$ for P2, also in agreement 
with  \eqref{eq:massesP1P2} for the eigenvalues of 
the positive-definite mass-matrix squared.

Hence, the difference in the neutralino masses and mixing matrix between both paths is beyond the precision of our computation of electroweak Sudakov logarithms.\footnote{But not for the Somerfeld effect, which is sensitive to the precise mass differences of the physical neutralino states in the EW broken theory.} Even though it looks more natural in our computation to follow path 1, the code developed for computing Sommerfeld enhancements in \cite{Beneke:2012tg} follows path 2, and so path 2 is used in the numerical analysis. 

\section{Summary of factorization functions}
\label{app:FunctionsDef}

This appendix collects the precise definitions of different functions appearing in the factorization theorem \eqref{def:GammaIJ}. This work generalizes the case of a pure SU(2) multiplet  described in great detail in \cite{Beneke:2019vhz} and extended to the pure Higgsino model in \cite{Beneke:2019gtg} and largely follows their notation.

\subsection{The hard function  $H_{ij}$}
The physics of the hard scale is entirely captured by the NREFT Wilson coefficients $C_i(\mu)$ for the annihilation operator basis (\ref{def:ops}). The two annihilation vertices appearing in the computation of the semi-inclusive cross section do not need to be identical and the hard function is given as a product of two Wilson coefficients
\begin{equation}
H_{ij}(\mu_h,\mu) \, \equiv \, C_i(\mu) \,C_j^\ast(\mu)\,.
\end{equation}
The natural scale of the hard function $H_{ij}$ is 
$\mu_h \sim 2\mDM$. It is evolved to the virtuality scale $\mu$ 
by solving the RGE \eqref{eq:hardRGE} with the anomalous dimensions given in section \ref{sec:ADM}.

\subsection{The photon function  $Z_\gamma^{WY}$}
At LO the unresummed photon function is simply
\begin{equation}
\left. Z_\gamma^{YW}(\mu_\gamma,\nu_h,\mu,\nu)\right|_{LO} = \left(
\begin{matrix}
s_W^2(\mu) & s_W(\mu)\,c_W(\mu) \\
s_W(\mu)\,c_W(\mu) & c_W^2(\mu)
\end{matrix}
\right)^{\!YW}
\quad \text{with}\,\, Y,W\,=\,3,4\,,
\end{equation}
which follows from the projection of the emitted SU(2) or U(1)$_Y$ gauge bosons onto the external photon state.\footnote{There is a sign difference between our $Z_\gamma^{34}$, $Z_\gamma^{43}$ and (C.15) of \cite{Beneke:2019gtg} due to a different assignment of electromagnetic charge states to the components of $\chi$. } Here the Weinberg angle 
is defined in the $\overline{\rm MS}$ scheme in terms of the SU(2) and U(1)$_Y$ gauge couplings.
The running is particularly simple in our resummation scheme since $\mu_s = \mu_\gamma=m_W$, where the photon function rapidity evolution factors are trivial $V_\gamma^{YW}(m_W,\nu,\nu_f)=1$, as can be seen in eqs.~(3.22) of \cite{Beneke:2019vhz} (for $Z^{33}_\gamma$) and (C.23) of \cite{Beneke:2019gtg}. It is still necessary to include rapidity resummation in the presence of scale variations $m_W \sim \mu_\gamma=\mu_s \neq m_W$.

\subsection{The jet function  $J_{\text{int}}^{\mathcal{G}}$}
The jet function for intermediate photon-energy resolution 
$E^\gamma_{\rm res}$ is split into the two separate gauge group parts
$J_{\text{int}}^{\text{U(1)}}(\mu_j,\mu)$ and $J_{\text{int}}^{\text{SU(2)}}(\mu_j,\mu)$. Both unresummed jet functions are trivial at LO 
\begin{equation}
\left. J_{\text{int}}^{U(1),\,SU(2)}(p^2,\,\mu_j,\mu)\right|_{LO}=\delta(p^2)\,.
\end{equation}
In order to write the scale evolution as a multiplicative factor $U_{U(1),SU(2)}$, we  transform the non-abelian jet function to Laplace space $j_{\text{int}}^{\text{SU(2)}}$. 
Complete NLO expressions for the unresummed jet functions are given in eq.~(3.27) of \cite{Beneke:2019vhz} and eq.~(C.24) of \cite{Beneke:2019gtg}. At the NLL order, we only need
\begin{align}
\left. J^{\text{U(1)}}_{\text{int}} (p^2; \mu_j,\mu)\right|_{NLL} &= \left. U_J^{\text{U(1)}}(\mu_j,\mu)  \, J^{\text{U(1)}}_{\text{int}} (p^2, \mu_j,\mu_j)\right|_{LO} = U_J^{\text{U(1)}}(\mu_j,\mu) \,\delta(p^2)\,, \\
\notag
\left. J^{\text{SU(2)}}_{\text{int}} (p^2, \mu_j,\mu) \right|_{NLL}  &= \left. U_J^{\text{SU(2)}}(\mu_j,\mu)  \, j^{\text{SU(2)}}_{\text{int}} (\partial_\eta; \mu_j,\mu_j)\right|_{LO} \frac{e^{-\gamma_E \eta}}{\Gamma(\eta)}\frac{1}{p^2}\left( \frac{p^2}{\mu_j^2}\right)^\eta =\\
& = U_J^{\text{SU(2)}}(\mu_j,\mu) \,\frac{e^{-\gamma_E \eta}}{\Gamma(\eta)}\,\frac{(p^2)^{\eta-1}}{(\mu_j^2)^\eta}.
\end{align}

\subsection{The soft function  $W_{IJ,WY}^{\mathcal{G},\,ij}$}
\label{app:soft}
Suppressing the dependence on the resummation scales, since we choose them at the soft scale $\mu=\mu_s$ and $\nu=\nu_s$, the tree-level soft function is given by
\begin{align}\label{def:TLsoft}
\left.  W^{\text{SU(2)}\,ij}_{IJ,WY}(\omega) \right|_{LO} &\,=\, \delta(\omega) \cdot w^{\text{SU(2)}\,ij}_{IJ,WY} \,=\, \delta(\omega) \cdot (\mathcal{S}^j_{J,3Y})^\ast \, (\mathcal{S}^i_{I,3W})\,,
\nn \\[5pt]
\left.  W^{\text{U(1)}\,ij}_{IJ,WY}(\omega) \right|_{LO} &\,=\,\, \delta(\omega) \cdot w^{\text{U(1)}\,ij}_{IJ,WY} \,\,=\, \delta(\omega) \cdot (\mathcal{S}^j_{J,4Y})^\ast \, (\mathcal{S}^i_{I,4W})\,,
\end{align}
where
\begin{align}
\mathcal{S}^i_{I,VW}=K_{ab,I}(T^{VW}_i)_{ab}\,,
\end{align}
which is derived from eq.~(2.51) in \cite{Beneke:2019vhz} after expanding the Wilson lines  $Y_v\approx 1$ at LO and generalizing the adjoint indices $A,B$ of SU(2) to SU(2)$\otimes$U(1)$_Y$. 
The factor $T^{VW}_i$ can be extracted from \eqref{def:genOp} for each operator of the operator basis \eqref{def:ops}. The tensor $K_{ab,I}$ transforms fermion bilinears from the gauge basis $a$, $b$ to the mass eigenstate basis $I=(i_1 i_2)$ after EWSB, thus 
\begin{align}\label{def:K}
K_{ab,I}=(RG)^{-1}_{ai_1}(RG)^{-1}_{bi_2}
\end{align}
in terms of the matrices $G$, $R$ defined in appendix~\ref{app:EWSB}.
It is not possible to give $W_{IJ,WY}$ in a reasonably compact, more explicit form than \eqref{def:K} since $GR$ depends on the specific MSSM model parameters.  

The running of the soft function is obtained by requiring (by consistency) that the final result is independent of the resummation scales $\mu,\nu$:
\begin{align}
\frac{d\Gamma_{IJ}}{d\ln\mu}=\frac{d\Gamma_{IJ}}{d\ln\nu}=0.
\end{align}
We refer to section 3.2 and appendix C of \cite{Beneke:2019vhz} for details on the simultaneous rapidity and virtuality regularization. 


\section{Computation of the anomalous dimension}
\label{app:ADM}

The contributions from Higgs-Higgsino-gaugino interaction depicted in figure~\ref{fig:1loop} cannot be inferred from standard gauge-theory results \cite{Beneke:2009rj} and need to be computed explicitly. They give rise to the $a_+^2$ contributions in \eqref{GammaAA}--\eqref{GammaBB}. 

The appearances of $a_+^2$ in the diagonal entries of the anomalous dimension matrices arise from Higgs loop contributions to the field renormalization constants of gauginos and Higgsinos, as depicted on the left of figure~\ref{fig:1loop}. The  $\overline{\rm MS}$ field renormalization constants now read
in dimensional regularization $d=4-2\epsilon$:
\begin{align}
Z_{\chi_A}&=1+\frac{g_2^2}{(4\pi)^2}\frac{4}{\epsilon}-\frac{g_2^2}{(4\pi)^2}\frac{2a_+^2}{\epsilon}\,,\\
Z_{\chi_H}&=1+\frac{g_2^2}{(4\pi)^2}\frac{3}{2\epsilon}+\frac{g_1^2}{(4\pi)^2}\frac{1}{2\epsilon}-\frac{g_1^2+3g_2^2}{(4\pi)^2}\frac{a_+^2}{2\epsilon}\,,\\
Z_{\chi_B}&=1-\frac{g_1^2}{(4\pi)^2}\frac{2a_+^2}{\epsilon}.
\end{align}
Apart from the entry $[\gamma_{AA}]_{21}$, off-diagonal contributions to \eqref{GammaAA}--\eqref{GammaBB} originate from operator mixing from vertex diagrams as in the right of figure~\ref{fig:1loop}.
As an illustrative example we give here the explicit expression for the mixing of operators $\mathcal{O}_3$ and $\mathcal{O}_4$ into each other ($v$ is the heavy-particle four-velocity satisfying $v^2=1$):
\begin{align}
\langle \mathcal{O}_3 \rangle^\text{bare}_\text{1-loop} &= 
a_+^2\left( -i \frac{g_1}{2}\delta^{li} \right)
\left( i \frac{g_1}{2}\delta^{lj} \right)
2\, \eta^\dagger_{Hi}\Gamma^{\mu\nu} \xi_{Hj} \, \epsilon^*_\mu(n) \, \epsilon^*_\nu(\bar{n}) \,\mu^{2\epsilon} \nn\\
&\times \int\frac{d^dl}{(2\pi)^d}\frac{-i}{(v\cdot l+i\epsilon)(v\cdot l-i\epsilon)(l^2+i\epsilon)}
=
\frac{-g_1^2}{16\pi^2}\frac{a_+^2}{2\epsilon}\langle \mathcal{O}_4 \rangle_\text{tree}+\mathcal{O}(\epsilon^0)\,,\label{Z34}\\[5pt]
\langle \mathcal{O}_4 \rangle^\text{bare}_\text{1-loop} &= 
2a_+^2\left( -i \frac{g_1}{2}\delta^{li} \right) \delta^{ij} \left( i \frac{g_1}{2}\delta^{lj} \right) 
2\, \xi^{c \dagger} \Gamma^{\mu\nu} \xi \, \epsilon^*_\mu(n) \, \epsilon^*_\nu(\bar{n})\,\mu^{2\epsilon}  \nn\\
&\times \int\frac{d^dl}{(2\pi)^d}\frac{-i}{(v\cdot l+i\epsilon)(v\cdot l-i\epsilon)(l^2+i\epsilon)}
=
\frac{-g_1^2}{16\pi^2}\frac{2a_+^2}{\epsilon}\langle \mathcal{O}_3 \rangle_\text{tree}+\mathcal{O}(\epsilon^0). \label{Z43}
\end{align}
The masses $M_{h,H}$ in the Higgs propagator can be dropped for the computation of the divergent part, thus one finds a factor $a_+^2$ which arises from decomposing $ H_+$ in terms of mass-eigenstate fields. Note the additional symmetry factor of 2 in \eqref{Z43} with respect to \eqref{Z34} that originates from the indistinguishable external Majorana fermions.


\section{Matching the anomalous dimension for large mass splittings}
\label{app:Matching}

\begin{figure}[t]
\begin{center}
\includegraphics[width=0.6\textwidth]{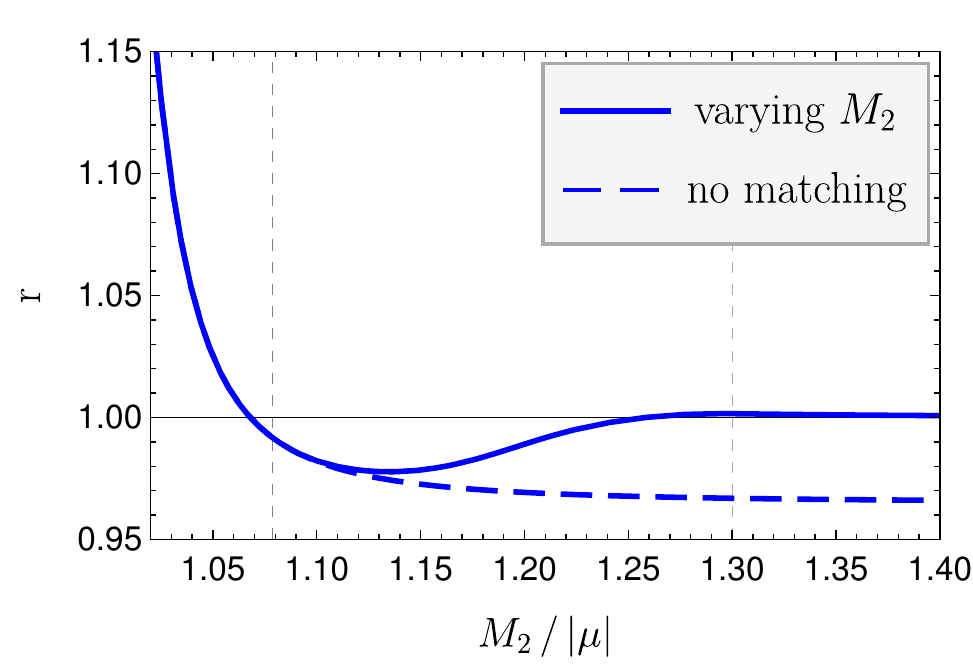}
\end{center}
\caption{Ratio $r$ of the diagonal lightest-chargino annihilation matrix element in $\Gamma_{IJ}^{NLL,imp}$ to the heavy-wino limit (approximated by $M_2=4 M_1$) when varying $M_2$ in benchmark "BHW-mix".
}
\label{fig:decoupling}
\end{figure}

When some neutralino mass splittings become significantly larger than $\mathcal{O}(m_W)$, the heavier neutralino states should be removed from the non-relativistic Lagrangian, and their contribution to the anomalous dimensions in section~\ref{sec:Ann} should be decoupled, since the loops with heavy states do not cause large logarithms. The decoupling is not automatic for diagrams involving the Higgs-Higgsino-gaugino interaction. As an example, consider the bino one-loop self-energy correction with internal Higgsino and Higgs propagators. Taking $n_H=2$, but neglecting the internal Higgs masses (which also cut off the logarithmic enhancement whenever they are larger than the Higgsino-bino mass splitting) and defining the mass ratio of internal and external particle masses $x\equiv\frac{|\mu|}{M_1}$, one finds 
\begin{equation}\label{winoSE}
\frac{\partial\Sigma}{\partial\slashed{p}}\Bigg |_{\slashed{p}=M_1,\,\text{IR}}=\frac{g_1^2}{16\pi^2}\left(2+4x^2+4x^4\ln(1-\frac{1}{x^2})\right)\,.
\end{equation}
Clearly, the logarithmic enhancement vanishes when 
$x$ is not close to 1. Since this feature is not reproduced 
in the dimensionally-regulated NREFT, in order to correctly incorporate the logarithms only at small mass-splittings we introduce a decoupling function that switches off the specific contributions from the neutralino-species changing Higgs-loop diagrams to the diagonal and off-diagonal hard anomalous dimension matrix elements when mass differences become large. We invoke an arc-tangent for a smooth decoupling function $f:x\mapsto [0,1]$, defined by 
\begin{equation}
f(x) = \left\lbrace
\begin{array}{l}
f(x)=1,\quad x<t_1,\\
f(x)=0,\quad x>t_2,\\
\frac{1}{2}- \frac{1}{\pi}\arctan\left[\left(\frac{-1}{|t_1-x|^2}+\frac{1}{|t_2-x|^2}\right)\cdot \frac{\pi}{4}\,\left|\frac{t_2-t_1}{2}\right|^2\right], \quad\mbox{else}.
\end{array}
\right.
\label{MatchingFn}
\end{equation}
The interpolation takes place between the thresholds $t_1<t_2$ in the mass ratio $x$. The mass differences in \eqref{LagNRg} are counted as $\mathcal{O}(m_W)$, therefore we choose $t_1=1+\frac{2m_W}{m_\text{LSP}}$. The value of $t_2$ is taken to be $t_2=1.3$, allowing 
for a 30\% mass non-degeneracy before the anomalous 
dimension from the heavy particle is turned off. 
This allows for a simple implementation in the numerical code, since the Higgsino-gaugino contributions are proportional to $a_+^2$. Hence we replace $a_+^2\to a_+^2\cdot f(x)$ with $x$ adapted to the ratio of  internal and external particle masses in each contribution, which are uniquely identified by the power of the gauge couplings $g_{1,2}$ in the interaction. In \eqref{def:GammaImp}, $\Gamma_{IJ}^{NLL, \rm imp}$ implicitly includes the decoupling function $f(x)$ in addition to the improvement factor $\mathcal{G}_{IJ}$ discussed in the main text, such that the limit of large mass splittings is correctly described. 

Figure \ref{fig:decoupling} illustrates the the decoupling for the lightest-chargino channel by using benchmark "BHW-mix" (for which  $t_1\approx 1.08$) as a starting point from which $M_2$ is gradually increased while $M_1$ and $\mu$ are kept fixed. The bold and dashed lines show the ratio $r$ with and without ($f(x)=1$) the matching implemented, respectively, normalized to the heavy-wino limit $M_2 \gg \mDM$ for the diagonal lightest-chargino annihilation matrix element $I=J=\chi_1^+\chi_1^-$:
\begin{equation}\label{eq:DeltaGamma}
r_{IJ} \equiv \frac{ \Gamma_{IJ}^{NLL,\rm imp} }{ \Gamma_{IJ}^{NLL,\rm imp} |_{M_2\gg\mDM}}, \qquad
r_{IJ}^{(no\,matching)} \equiv  \frac{ \Gamma_{IJ}^{NLL,\rm imp} |_{f(x)=1}}{ \Gamma_{IJ}^{NLL,\rm imp} |_{M_2\gg\mDM}}  \,.
\end{equation}
The decoupling at large $x$ is described correctly as $r_{IJ}\to 1$ for $x\to \infty$, while the dashed line without decoupling approaches the incorrect value $0.96$.


\section{Benchmark data}
\label{app:data}


\begin{table}[p]\begin{center}
\begin{tabular}{r|cccccc}
model&$\mu$&$M_1$&$M_2$&$\tan\beta$&$M_A$&$m_{sf}$
\\
\hline
\emph{pure models}\quad\quad\quad&&&&
\\
\hline
\hline
H&-1.112&7.5&7&15&2.9&12\\
W&9&8&2.85&15&2.9&13\\[3pt]
\emph{2-component}\quad\quad&&&&
\\
\hline
\hline
BH&-1.069&1.116&10&2.24&2.9&29\\
BW&2.91&2.145&2.148&15&2.9&12\\
BW-e&-2.27&1.829&1.836&15&2.446&5.45\\
BW-2520&-3.3&2.525&2.52&20&2.5&7.25\\
BW-e-nh2&3.6&2.055&2.058&15&0.55&10\\
HW&-2.977&30&2.839&3&2.9&20\\
HW-nh2&-3.065&20&2.92&3&0.78&25\\[3pt]
\emph{fully mixed}\quad\quad\quad&&&&
\\
\hline
\hline
BHW-mix&-2.041&1.92&1.929&2.5&1.96&20\\
BHW-mix2&-2.094&1.97&1.978&2.5&2.045&20\\
BHW-mass&-1.701&1.625&1.642&2.2&1.725&25\\
BHW-nh2&-1.94&1.802&1.819&3.6&0.67&25\\[3pt]
\emph{additional}\quad\quad\quad&&&&&&
\\
\hline
\hline
$^\ast$B&7&2.145&8&15&2.9&12\\
$^\ast$BW-coan&7&2.145&2.148&15&2.9&12\\
$^\ast$H+&1.112&7.5&7&15&2.9&12\\
H2&-1.24&8&1.419&2.4&1.25&29\\
BH-undet&-1.363&1.3&1.33&2.19&1.305&25\\
BW-nfw&3.38&2.075&2.078&15&2.9&12\\
BW-ce&-3.2&2.287&2.288&13&2.8&13.8\\
BW-2670&-3.1&2.677&2.67&20&2.7&12\\
BW-nh2&-3.15&2.4417&2.44&15&0.62&9\\
BW-ce-nh2&3.42&2.165&2.1665&15&0.57&9.5
\end{tabular}
\caption{\label{tab:DataInput}
Numerical inputs for used models. Masses are given in TeV. An asterisk $\ast$ indicates that this model does not produce the observed DM relic density.}
\end{center}\end{table}


We consider CP-preserving and flavour-diagonal MSSM models. The gluino 
mass is irrelevant for this calculation and we set it to 
$m_3=30\,\text{GeV}$. The sfermion masses are assumed to be large enough to make 
the sfermions essentially decoupled. In practice, we set all 
sfermion masses to a common mass $m_{sf} > 2\mDM$, in most models even above $4\mDM$. This leaves five MSSM 
parameters: the bino, wino and Higgsino mass parametes $M_1$, $M_2$, $\mu$, respectively, as well as the Higgs vev-ratio $\tan\beta$ and the 
CP-odd Higgs mass $M_A$.  
We used \texttt{FeynHiggs} \cite{Bahl:2018qog} 
(version 2.14.4) 
to generate the MSSM model parameters. The details for the 
23 models analyzed in this work are shown in table~\ref{tab:DataInput}.


\newpage

\providecommand{\href}[2]{#2}\begingroup\raggedright\endgroup


\end{document}